\pdfoutput=1

\documentclass[11pt]{article}

\usepackage[preprint]{acl}

\usepackage{times}
\usepackage{latexsym}
\usepackage{fontawesome}

\usepackage[T1]{fontenc}

\usepackage[utf8]{inputenc}

\usepackage{microtype}

\usepackage{inconsolata}

\definecolor{cvprblue}{rgb}{0.21,0.49,0.74}

\usepackage{multirow}
\usepackage{CJKutf8}
\usepackage{adjustbox}
\usepackage{tabularx}
\usepackage{rotating}
\usepackage{amsmath}
\usepackage{amsfonts} 
\usepackage{amssymb} 
\usepackage{algorithm}
\usepackage{algorithmic}

\usepackage{multicol} 
\usepackage{caption} 
\definecolor{myred}{RGB}{255,0,0}
\definecolor{myyellow}{RGB}{255,255,0}
\definecolor{mygreen}{RGB}{0,128,0}
\definecolor{myblue}{RGB}{0,0,255}
\definecolor{mygray}{RGB}{128,128,128}
\usepackage{booktabs}
\definecolor{deeppurple}{RGB}{102, 0, 153}

\newcommand{\SX}[1]{\textcolor{black}{#1}} 
\newcommand{\XS}[1]{\textcolor{black}{#1}} 

\title{Diverse Sign Language Translation}

\author{
\parbox{1\linewidth}{\centering
Xin Shen$^1$, Lei Shen$^2$, Shaozu Yuan$^2$, Heming Du$^1$, Haiyang Sun$^3$, Xin Yu$^1$}
\\
The University of Queensland, Brisbane, Australia\\
JD AI Research, Beijing, China \\
Li Auto Inc., Beijing, China
\\
\texttt{x.shen3@uqconnect.edu.au}}

\begin{document}
\maketitle
\begin{abstract}
Like spoken languages, a single sign language expression could correspond to multiple valid textual interpretations. 
Hence, learning a rigid one-to-one mapping for sign language translation (SLT) models might be inadequate, particularly in the case of limited data. 
In this work, we introduce a Diverse Sign Language Translation (DivSLT) task, aiming to generate diverse yet accurate translations for sign language videos. 
Firstly, we employ large language models (LLM) to generate multiple references for the widely-used CSL-Daily and PHOENIX14T SLT datasets. Here, native speakers are only invited to touch up inaccurate references, thus significantly improving the annotation efficiency.
Secondly, we provide a benchmark model to spur research in this task. 
Specifically, we investigate multi-reference training strategies to enable our DivSLT model to achieve diverse translations.
Then, to enhance translation accuracy, we employ the max-reward-driven reinforcement learning objective that maximizes the reward of the translated result. 
\SX{Additionally, we utilize multiple metrics to assess the accuracy, diversity, and semantic precision of the DivSLT task.}
Experimental results on the enriched datasets demonstrate that our DivSLT method achieves not only better translation performance but also diverse translation results.
\end{abstract}

\begin{figure}[t]
\begin{center}
   \includegraphics[width=1\linewidth]{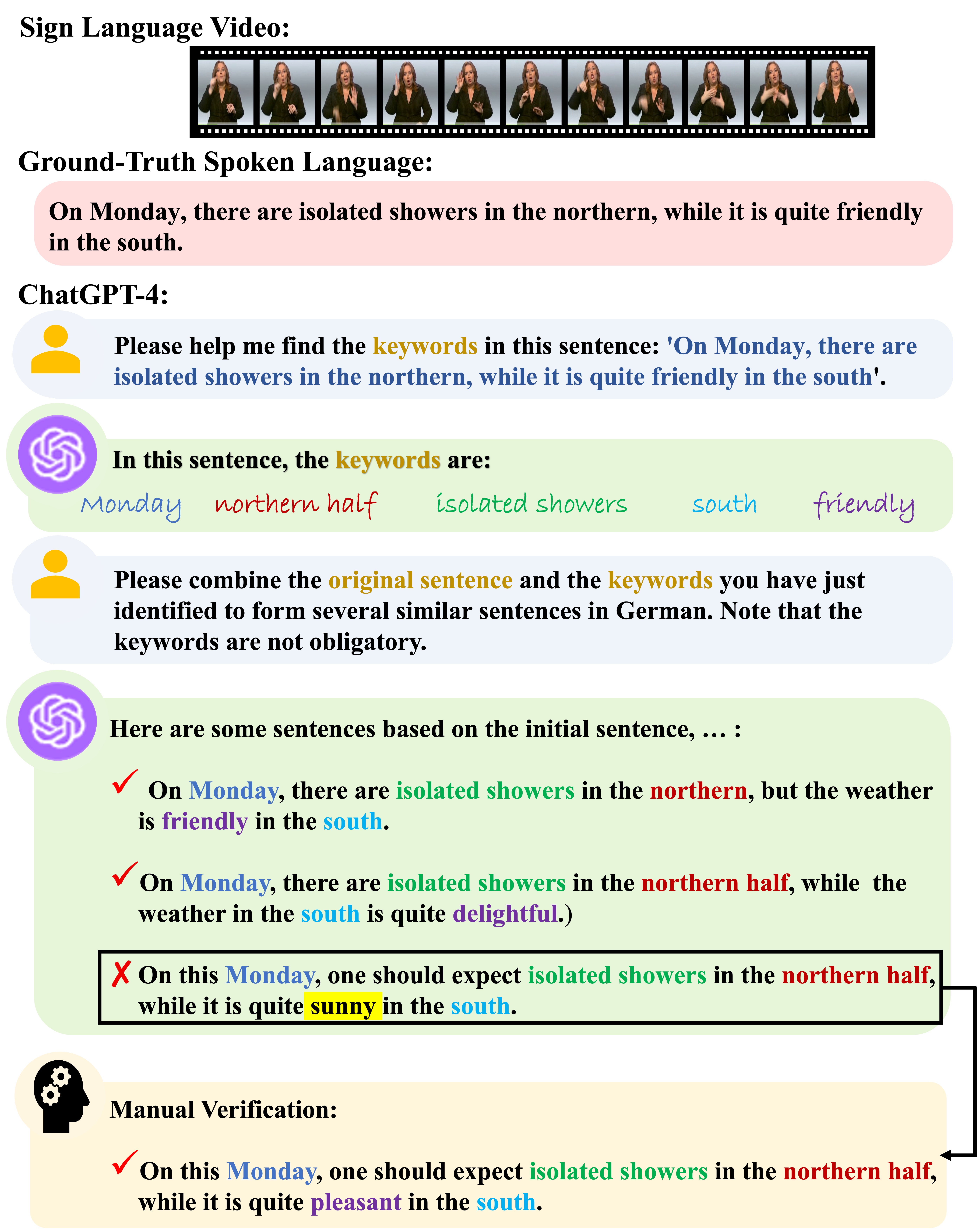}
   \vspace{-1.3em}
   \caption{Illustration of our pipeline of leveraging an LLM to generate multiple translations that closely resemble the ground-truth translation. 
   }
\label{intro_fig}
\end{center}
\vspace{-1em}
\end{figure}

\section{Introduction}
\label{sec:intro}

Sign language translation (SLT) has attracted great attention from the research field of computer vision and natural language processing ~\cite{NSLT,CSL_Daily,how2sign,BOBSL,KSL,shen2023auslan,gloss_free_m1,gloss_free_m2,GF_VLP,iclr23}.
Despite the remarkable achievement of progress, SLT models still face the following challenges:
(1) in the situation of data sparsity, an emphasis on pursuing translation accuracy during training may lead to a bias in SLT models towards common phrases~\cite{multi_ref_improve_NMT, Multi_Candidate_Opt}; 
(2) variability and expressiveness are also the characteristics of sign languages~\cite{variability&expressiveness}, and sign videos should be permitted to be translated into different expressions with similar semantics.

As illustrated in Figure~\ref{intro_fig}, \textit{``it is quite friendly in the south''} and \textit{``the weather in the south is quite delightful''} have different sentence structures but convey the same meaning.
However, existing SLT models will receive a penalty if their translation results for a sign video do not resemble the single ground-truth provided in the current SLT datasets. Thus, SLT models might fail to generate high-quality diverse translations. 
Actually, diverse translations can also potentially facilitate individuals to better understand sign language, especially when encountering sign words or expressions with multiple meanings (polysemy)~\cite{wlasl,sheng2024ai}. 
Driven by this, we argue that providing multiple translations for a sign video is more reasonable and preferable for training an SLT model. 
Hence, this work aims to tackle the Diverse Sign Language Translation (DivSLT) task.

To validate the study of DivSLT, we enrich the existing SLT datasets so that each sign video encompasses multiple references.
Without a doubt, it would be tedious to provide various translations of a sign video by humans.
Instead, we opt to leverage large language models (LLM)~\cite{openai2023chatgpt,glm_130b,llama2} to assist the annotation process.
To be specific, we choose two widely-used SLT datasets as the data source, \textit{i.e.}, CSL-Daily (Chinese Sign Language $\leftrightarrow$ Chinese)~\cite{CSL_Daily} and PHOENIX14T (German Sign Language $\leftrightarrow$ German)~\cite{NSLT}. 
As shown in Figure~\ref{intro_fig}, for each sign video, we first employ ChatGPT-4~\cite{openai2023chatgpt} to extract keywords from the original reference, and then prompt it to generate multiple faithful translations using the original reference along with the extracted keywords.
Finally, to ensure reference accuracy, native speakers are invited to review and modify the generated texts.

To facilitate further research on the DivSLT task, we introduce a simple and effective two-stage training paradigm as the benchmark.
In the first stage, we enforce our SLT model to generate diverse translations with the supervision of the multi-references.
For each sign video, we construct a training pair by selecting a reference from the multi-reference set and each reference has its corresponding video.
In this way, we not only expand the training data but also enable the SLT model to acquire various translations.
Since each video corresponds to multiple references, this could cause ambiguity in learning.

In the second stage, to \SX{further} enhance the accuracy of diversified translation results, we employ a max-reward-driven reinforcement learning~\cite{RL1992} objective that maximizes the reward of a translated result. 
Here, we employ a sentence-level evaluation score, \textit{i.e.}, Bilingual Evaluation Understudy~\cite{bleu} (BLEU)\footnote{All BLEU scores mentioned in this paper refer to~\href{https://github.com/Maluuba/nlg-eval}{Corpus BLEU}. Unless otherwise specified, BLEU refers to BLEU-4.}, as the reward to finetune our DivSLT model.
Different from the word-to-word cross-entropy loss commonly used in SLT, reinforcement learning can directly optimize the accuracy of the translated sentences, thus leading to better translation performance~\cite{One_Ref_Not_Enough, video_caption_RL_copy, mixed_RL_2}. 
\SX{We utilize multiple applicable evaluation metrics to assess the DivSLT task, conduct extensive experiments on the DivSLT datasets and report results in terms of translation accuracy, diversity and semantic precision.} 
It is observed that our method achieves superior translation performance and produces more diverse translation results compared with one-to-one SLT models.
Overall, the contributions of our work are threefold:
(1) To the best of our knowledge, our work is the first attempt to address the diverse sign language translation (DivSLT) task that aims to achieve accurate yet diverse translation results. 
(2) We effectively extend two widely used SLT datasets, \textit{i.e.}, CSL-Daily and PHOENIX14T, with additional multi-reference annotations, named CSL-Daily-Div and PHOENIX14T-Div. Our newly annotated datasets offer an ideal testbed for the DivSLT task.
(3) To assist in research on the DivSLT task, we provide a two-stage training paradigm method as a strong baseline model and suitable evaluation metrics. The codes and new annotations will be publicly available for reproducibility~\href{https://github.com/XinS0909/Diverse_Sign_Language_Translation}{\faGithub~\textcolor{deeppurple}{DivSLT}}.
\vspace{-0.2em}
\section{Related Work}
\label{sec:related_work}
\vspace{-0.2em}
\subsection{Sign Language Translation}
\vspace{-0.2em}
Sign Language Translation (SLT)~\cite{NSLT} aims to convert a continuous sign language video clip into a spoken language. 
Current SLT models can be divided into three categories: Sign2Gloss2Text (S2G2T)~\cite{yin2020better} , Sign2<Gloss+Text> (S2<G+T>)~\cite{sltt,zhou2021spatial,mmtlb, two_s}, and Sign2Text (S2T)~\cite{tspnet, gloss_free_m1, gloss_free_m2,gloss_free_m9,GF_VLP, e2e_pose_SLT}.
The S2G2T model involve two stages: Sign2Gloss (S2G)~\cite{two_s, cslr1, cslr2, cslr3} and Gloss2Text (G2T)~\cite{iclr23, g2t1}.
Both S2<G+T> and S2T are end-to-end models. 
Compared to S2T models, S2<G+T> models use gloss information as auxiliary supervision for video feature extraction, thereby improving translation results. 
However, gloss annotations require sophisticated expertise and are costly and time-consuming~\cite{sign_align}.
Thus, it is hard to learn S2<G+T> and S2G2T models from datasets without glosses.
In this work, we focus on S2T (gloss-free) SLT.

\vspace{-0.2em}
\subsection{Diverse Neural Machine Translation}
\vspace{-0.2em}
Improvement of translation diversity is a popular topic in neural machine translation (NMT)~\cite{DivNMT_1,DivNMT_2,DivNMT_3, DivNMT_4}. 
Existing diverse NMT works can be categorized into three groups. 
The first group produces diverse translations by applying diversity regularization to decoding algorithms~\cite{li2016simple, vijayakumar2018}. 
The second one improves diversity by sampling from a mixture of models~\cite{shen2019, wu2020},
while the third group~\cite{shu2019, LachauxJL2020,sun2020} provides diversity signals to the decoding procedure. 
Despite their success, these approaches face the issues caused by the discrepancy between training and inference, that is, learning diverse translations from a single-reference corpus. 
\citet{Multi_Candidate_Opt} argue that such discrepancy prevents the models from learning one-to-many relations efficiently. 
Therefore, constructing a dataset with multiple references is critical to enhance the translation diversity.
In this work, we annotate the widely-used sign language datasets with additional diverse and high-quality references.

\subsection{Multimodal Text Generation}
Multimodal text generation \cite{vc_1,video_caption_RL_copy,nlg_1,image_caption_sample_one, mmnlg1, mmnlg2, mmnlg3} aims to generate corresponding text for the given multimodal data, including images, videos, and audio signals.
SLT requires a model to understand the gestures and intricate finger movements of signers and represent them in a spoken language. Additionally, the translation should be unaffected by environment changes. Moreover, as SLT is designed to facilitate communication, it should accommodate various translations while maintaining the same gist.
Hence, it is difficult to apply multimodal text generation directly to DivSLT, especially when data is limited and the translation space is large.


\subsection{Reinforcement Learning in NMT}
Reinforcement learning (RL) has recently drawn great attention in NMT, as it allows the optimization of non-differentiable objectives and reduces exposure bias in auto-regressive sequence generators. 
The works~\cite{RLNMT_1, RLNMT_2, RLNMT_3} employ the reinforce algorithm~\cite{RL1992} to optimize models with metric-based reward for text-based machine translation.
\citet{RLNMT_4} propose two reverse NMT models through a dual learning mechanism while ~\citet{RLNMT_5} employ a critic network to predict the reward.
Driven by these works, we intend to explore how high-quality diverse sign translation can be achieved via RL in this work.

\section{Diverse SLT Dataset Curation}
\label{sec:dataset}
\subsection{Multi-Reference Generation}
Large Language Models (LLM)~\cite{openai2023chatgpt,glm_130b,llama2} have demonstrated their superior abilities in various text generation tasks, \textit{e.g.}, machine translation~\cite{chatgpt_good_translator}, summarization~\cite{llm_summary} and dialogue~\cite{openai2023chatgpt}. 
Here, we employ LLM to assist in generating sign language translations that are both diverse and semantically congruent with the ground-truth spoken language.
To obtain superior multi-reference and minimize the requirement for extensive manual modifications, we conduct evaluations across a range of LLMs\footnote{Analyses of generation outcomes across different LLMs are provided in the Appendix.}.
Based on the evaluation, we utilize ChatGPT-4 to generate multi-references in this work.
\begin{table}[t]
\centering
\footnotesize
\caption{Human evaluation of reference generation with different prompts. 
Scoring Criteria: ``2'' - Semantically equivalent with diverse content; ``1'' - Semantically equivalent, content similar to input; ``0'' - Semantically different from input.
EV and Kw indicate the evaluator and keywords, respectively.}
\vspace{-1em}
\label{prompts}
\begin{tabular}{l|ccc|c}
\toprule
\multicolumn{1}{c|}{Prompt Type} & EV 1 & EV 2 & EV 3 & Avg. Score\\
\midrule
w/o Kw & 0.88 & 0.52 & 0.44 & 0.61 \\
w/ mandatory Kw & 1.24 & 1.06 & 1.10 & 1.13 \\
w/ optional Kw & 1.80 & 1.72 & 1.70 & \textbf{1.74} \\
\bottomrule
\end{tabular}
\end{table} 

\begin{table*}[!ht]
\caption{Comparison and statistics between the original and the extended multi-reference datasets. (OOV: out-of vocabulary, \textit{i.e.}, words occur in Dev set but not in Train set. Singleton: words that only occur once in Train set.)}
\vspace{-0.5em}
\label{statistics}
\tiny
\centering
\begin{tabularx}{0.9\textwidth}{l|cccccc|cccccc}
\toprule
 Dataset & \multicolumn{6}{c|}{PHOENIX14T~\cite{NSLT} (DGS $\leftrightarrow$ Germany)} & \multicolumn{6}{c}{CSL-Daily~\cite{CSL_Daily} (CSL $\leftrightarrow$ Chinese)} \\ \midrule
\multirow{2}{*}{split} & \multicolumn{2}{c}{Train} & \multicolumn{2}{c}{Dev} & \multicolumn{2}{c|}{Test} & \multicolumn{2}{c}{Train} & \multicolumn{2}{c}{Dev} & \multicolumn{2}{c}{Test} \\
 & original & extend & original & extend & original & extend & original & extend & original & extend & original & extend \\ \midrule
num.references & 7,096 & 42,576 & 519 & 3,114 & 642 & 3,852 & 18,401 & 110,406 & 1,077 & 6,462 & 1,176 & 7,056 \\
vocab. size & 2,887 & 9,427 & 951 & 2,619 & 1,001 & 2,893 & 2,343 & 2,951 & 1,358 & 1,899 & 1,358 & 1,899 \\
total words & 99K & 675K & 6.8K & 46.9K & 7.8K & 53.8K & 291K & 1.8M & 17K & 112K & 19K & 124K \\
total OOVs & - & - & 57 & 259 & 60 & 317 & - & - & 37 & 42 & 37 & 42 \\
singletons & 1,077 & 3,661 & - & - & - & - & 1 & 1 & - & - & - & - \\ \midrule
avg. pwb (k=5))~$\downarrow$ & - & 17.27 & - & 17.95 & - & 16.35 & - & 20.95 & - & 20.83 & - & 21.04 \\
avg. rfbRT (k=5)~$\uparrow$ & - & 70.08 & - & 71.07 & - & 70.46 & - & 66.89 & - & 67.35 & - & 67.24 
\\ \bottomrule
\end{tabularx} 
\vspace{-2em}
\end{table*}

Although ChatGPT-4 can provide multiple similar texts to the given ground-truth reference, we find that some of the generated sentences are not faithful to the original semantics.
To control the fine-grained content, we first employ ChatGPT-4 to extract keywords, and then generate multi-reference sentences based on the keywords and ground-truth reference.
Note that a sign video consists of a sequence of glosses and the spoken language reference is formed based on these glosses. 
It is observed that the keywords extracted by ChatGPT-4 are often aligned with glosses\footnote{In this work, gloss annotations are not used.}.

To generate multi-references with ChatGPT-4, we also investigate two types of prompts: the mandatory use of keywords, and the optional use of keywords.
We randomly choose 50 sign video clips\footnote{25 from CSL-Daily and 25 from PHOENIX14T.}, and then obtain generated results with these two types of prompts. Three native speakers are invited to score the quality of generated texts in terms of semantic similarity and diversity.
As shown in Table~\ref{prompts}, the references generated with the prompt type ``the optional use of keywords'' are most favored. 
Thus, we use this prompt to generate multi-references for each sign data.

\subsection{Dataset Statistics and Quality Evaluation}
To ensure the quality of the generated results, we invite native speakers to perform verification and modification. 
The modification rates for the Chinese and German DivSLT datasets are approximately 11\% and 7\%, respectively.
Our newly annotated diverse sign language translation dataset is named CSL-Daily-Div and PHOENIX14T-Div.
The statistics of the original and extended DivSLT datasets are presented in Table~\ref{statistics}.
We significantly increase the vocabulary size with the addition of more references, indicating considerable enhancement in the diversity of the extended datasets. 
However, there is a notable rise in out-of-vocabulary (OOV) words, particularly in German, which indicates challenges in DivSLT.

To measure the accuracy and diversity of multiple references in our DivSLT dataset, we use the following automatic evaluation metrics.
We denote the ground-truth reference in original dataset as $y$ and multiple generated references as \( \{y_1, \ldots, y_k\} \). 
The generation diversity is measured by the average pairwise BLEU (pwb):
\vspace{-0.8em}
\begin{equation}
\text{pwb} = \frac{1}{(k - 1)k} \sum_{i=1}^{k} \sum_{\substack{j \neq i}}^{k-1} \text{BLEU}(y_i, y_j),
\vspace{-0.6em}
\end{equation}
where lower pwb indicates better diversity.

Typically, reference BLEU (rfb)~\cite{One_Ref_Not_Enough} is used to measure the token-level accuracy, since the calculation is based on n-gram overlaps.
Therefore, there is a trade-off between accuracy rfb and pwb. 
Since we want to measure the semantic accuracy between sentences, BLEURT \cite{bleuRT} is a better choice.
BLEURT-20~\cite{bleuRT20} leverages BERT~\cite{bert} and is trained with rating data across 13 languages, encompassing both Chinese and German.
Therefore, following ~\citet{openASL} and \citet{e2e_pose_SLT}, we utilize BLEURT-20 as an automatic evaluation metric to measure the semantic similarity between generated references and the ground-truth reference (rfbRT):
\vspace{-0.5em}
\begin{equation}
\text{rfbRT} = \frac{1}{k} \sum_{i=1}^{k} \text{BLEURT}(y, y_i)\text{,}
\vspace{-0.5em}
\end{equation}
where higher rfbRT indicates better semantic accuracy. 
As shown in Table~\ref{statistics}, both PHOENIX14T-Div and CSL-Daily-Div endow high accuracy (higher rfbRT) and diversity (lower pwb). 
In other words, the additional generated references are suitable for the DivSLT task.
\section{Benchmark Method}
\label{sec:method}
\subsection{DivSLT Task Description}
Sign language translation is a typical sequence-to-sequence learning problem~\cite{NSLT}, which learns the conditional probability \( p(\mathbb{Y} | \mathbb{V}) \) of generating a spoken language sentence $\mathbb{Y}=\{y_m\}_{m=1}^M$ with $M$ words given a sign video clip $\mathbb{V}=\{v_t\}_{t=1}^T$ with $T$ frames.
In contrast, a DivSLT model aims to produce $K$ different candidate translations~$\mathcal{Y}=\{\mathbb{Y}_k\}_{k=1}^{K}$ given a sign video clip~$\mathbb{V}$. 
Similar to SLT, the DivSLT objective is also to minimize the cross-entropy loss:
\vspace{-0.5em}
\begin{equation}
L_{\text{CE}}(\theta) = - \sum_{n=1}^{N}\sum_{k=1}^{K} \log P_{\theta}(\mathbb{Y}_k^{(n)}|\mathbb{V}^{(n)})\text{,}
\vspace{-0.5em}
\end{equation}
where $\langle\mathbb{V}^{(n)}, \mathbb{Y}_k^{(n)}\rangle$ is the $n$-th sign video clip with its $k$-th reference in the training corpus of size $N$, and $P_{\theta}(\mathbb{Y}_k^{(n)}|\mathbb{V}^{(n)})$ denotes the estimated probability of the DivSLT model with parameters $\theta$. 

\subsection{DivSLT Model Architecture}
The gloss-free SLT models~(GFSLT)~\cite{GF_VLP, e2e_pose_SLT,gloss_free_m2} have demonstrated promising performance and even surpassed some gloss-based models in terms of efficacy. 
Without relying on additional information, \textit{i.e.}, gloss annotations, we utilize the vanilla gloss-free SLT in our experiments.
Motivated by~\citet{GF_VLP}, our DivSLT architecture is divided into two components: Visual Encoder and Text Decoder. 
Visual Encoder is designed to extract visual features. 
To be specific, ResNet~\cite{resnet} pre-trained on ImageNet~\cite{imagenet} is employed as the visual backbone. The temporal blocks~\cite{CSL_Daily} are utilized to capture the dynamics of sign gestures. 

Text Decoder is an encoder-decoder transformer, aimed to convert visual representations to textual output. 
Both the encoder and decoder have three layers, with a hidden size of 1024, a feed-forward layer size of 4096, and 8 attention heads in each layer. We initialize the decoder part from mBART-large-cc~\cite{mbart}.

\subsection{Two-stage DivSLT Training Paradigm}
\label{sec:Two-stage DivSLT}
\textbf{Stage 1: Multi-Reference Learning.} In the first stage, our primary goal is to enhance SLT model to generate diverse translations.
Our multi-reference training strategy is achieved by converting a multiple-reference dataset to its single-reference counterpart.
Considering a multi-reference dataset $D$, where the $n$-th training example $\langle\mathbb{V}^{(n)}, \mathcal{Y}^{(n)}\rangle$ includes one sign video $\mathbb{V}^{(n)}$ and a reference set $\mathcal{Y}^{(n)} = \{\mathbb{Y}_1^{(n)}, \mathbb{Y}_2^{(n)}, \ldots, \mathbb{Y}_K^{(n)}\}$ of $K$ references. 
As suggested by~\citet{multi_reference_train_method}, we construct the single reference dataset $D'$ from $D$ by duplicating the $\mathbb{V}^{(n)}$ sign video clip $K$ times, and each is accompanied by a distinct reference. This process is expressed as:
\vspace{-1em}
\begin{equation}
D' = \bigcup_{n=1}^{|D|}\bigcup_{k=1}^{K}{(\mathbb{V}^{(n)}, \mathbb{Y}_k^{(n)})}\text{.}
\vspace{-0.5em}
\end{equation}

Moreover, we found that randomly shuffling the order of the video-reference pairs can lead to better performance. This is because introducing randomness into the learning process helps the generalization ability of the learned network.
Afterward, we employ the cross-entropy loss to train our DivSLT model, as follows:
\vspace{-0.7em}
\begin{equation}
    L_{\text{CE}}(\theta) = - \sum_{n} \mathbb{Y}_{j,k}^{(n)} \log{\widehat{\mathbb{Y}_j}},
    \vspace{-1em}
\end{equation}
where $\widehat{\mathbb{Y}}$ is the word sequence sampled from our model, \( \theta \) represents the model parameters, and $j$ indicates the $j$-th word in the reference $\mathbb{Y}_k^{n}$.
{
\setlength{\tabcolsep}{5.5pt}
\begin{table*}[ht]
\centering
\tiny
\caption{\XS{Experimental results on the \textbf{test} set of PHOENIX14T-Div. ``BM'', ``MR'', ``R'', ``B'', and ``BRT'' represent best-matching, multi-reference, ROUGE, BLEU and BLEURT, respectively. $^*$ represents our reproduced results.}}
\vspace{-1em}
\label{DSLT_DE}
\begin{tabular}{@{}l|ccccc|cccccc@{}} \toprule
 & \multicolumn{5}{c|}{Top-3 Predictions} & \multicolumn{6}{c}{Top-1 Prediction} \\ \midrule
{\bf Single-Reference SLT} & rfb-BM~$\uparrow$ & mrfb~$\uparrow$ & \textbf{\textcolor{blue}{pwb}}~$\downarrow$ & rfbRT-BM~$\uparrow$ & mrfbRT~$\uparrow$ & BRT-MR~$\uparrow$ & R-MR~$\uparrow$ & B-MR~$\uparrow$ & \textbf{\textcolor{blue}{BRT-BM}}~$\uparrow$ & R-BM~$\uparrow$ & B-BM~$\uparrow$ \\ \midrule 
\XS{NSLT~\cite{NSLT}} & 10.05 & 12.87 & 48.20 & 38.42 & 30.99& 35.72 & 27.78 & 11.89  & 44.55 &31.79  & 12.77 \\
\XS{TSPNet-Joint~\cite{tspnet}} & 12.05&	15.24&	43.15&	42.22&	34.54& 40.15&	36.01&	18.61&	47.73&	35.66&	14.19 \\
GASLT~\cite{gloss_free_m2} & 15.56 & 17.49 & 43.01 & 45.91 & 38.10 & 41.57 & 37.94 & 18.66 & 49.07 & 36.54 & 16.45 \\
GFSLT~\cite{GF_VLP}$^*$  & 19.03 & 20.41 & 40.86 & 51.73 & 43.82  & 46.45 & 42.53 & 21.29 & 54.70 & 41.14 & 19.76 \\ \midrule \midrule
{\bf Multi-Reference SLT} & rfb-BM~$\uparrow$ & mrfb~$\uparrow$ & \textbf{\textcolor{blue}{pwb}}~$\downarrow$ & rfbRT-BM~$\uparrow$ & mrfbRT~$\uparrow$ & BRT-MR~$\uparrow$ & R-MR~$\uparrow$ & B-MR~$\uparrow$ & \textbf{\textcolor{blue}{BRT-BM}}~$\uparrow$ & R-BM~$\uparrow$ & B-BM~$\uparrow$ \\ \midrule 
DivSLT [Stage 1]  & 20.74 & 23.96 & \textbf{22.26} & \textbf{61.89} & 54.14 & 57.15 & 48.24 & 27.69 & 64.82&	46.02&	24.20   \\
DivSLT [Stage 2] & \textbf{22.12} & \textbf{25.23} & 26.30 & 61.34 & \textbf{54.97}  &  \textbf{58.02} & \textbf{49.82} & \textbf{28.82} & \textbf{64.95}&	\textbf{47.61}&	\textbf{25.29} \\ \bottomrule
\end{tabular}
\vspace{-1em}
\end{table*}
}
{
\setlength{\tabcolsep}{5.5pt}
\begin{table*}[ht]
\centering
\tiny
\caption{\XS{Experimental results on the \textbf{test} set of CSL-Daily-Div. 
``BM'', ``MR'', ``R'', ``B'', and ``BRT'' represent best-matching, multi-reference, ROUGE, BLEU and BLEURT, respectively. 
$^*$ represents our reproduced results.}}
\vspace{-1em}
\label{DSLT_ZH}
\begin{tabular}{@{}l|ccccc|cccccc@{}} \toprule
 & \multicolumn{5}{c|}{Top-3 Predictions} & \multicolumn{6}{c}{Top-1 Prediction}  \\ \midrule
\textbf{Single-Reference SLT} & rfb-BM~$\uparrow$ & mrfb~$\uparrow$ & \textbf{\textcolor{blue}{pwb}}~$\downarrow$ & rfbRT-BM~$\uparrow$ & mrfbRT~$\uparrow$ & BRT-MR~$\uparrow$ & R-MR~$\uparrow$ & B-MR~$\uparrow$ & \textbf{\textcolor{blue}{BRT-BM}}~$\uparrow$ & R-BM~$\uparrow$ & B-BM~$\uparrow$ \\ \midrule 
\XS{TSPNet-Joint~\cite{tspnet}$^*$}  & 2.93&	8.82&	47.89&	26.43&	24.29& 28.34&	20.69&	4.72&	33.13&	18.72&	3.23 \\
GASLT~\cite{gloss_free_m2}$^*$  & 5.56 & 8.90 & 47.51 & 27.44 & 24.88&29.09 & 24.81 & 6.35 & 34.94 & 24.25 & 5.74\\
\XS{NSLT~\cite{NSLT}}$^*$  & 8.18 & 10.41 & 43.79 & 30.08 & 25.35 & 30.79 & 32.71 & 9.89 & 36.55 &  31.24 & 7.96\\
GFSLT~\cite{GF_VLP}$^*$ & 9.75 & 11.31 & 39.47 & 31.96 & 26.17 & 31.51 & 34.23 & 11.67 & 37.19 & 30.85 & 8.94   \\ \midrule \midrule
\textbf{Multi-Reference SLT}& rfb-BM~$\uparrow$ & mrfb~$\uparrow$ & \textbf{\textcolor{blue}{pwb}}~$\downarrow$ & rfbRT-BM~$\uparrow$ & mrfbRT~$\uparrow$ & BRT-MR~$\uparrow$ & R-MR~$\uparrow$ & B-MR~$\uparrow$ & \textbf{\textcolor{blue}{BRT-BM}}~$\uparrow$ & R-BM~$\uparrow$ & B-BM~$\uparrow$ \\ \midrule 
DivSLT [Stage 1] & 11.88 & 13.98 & \textbf{22.65}  & 39.80 & 33.26 & 35.51 & 34.32 & 14.06 &  42.39 &	31.80 &	12.11  \\
DivSLT [Stage 2] & \textbf{13.24} & \textbf{15.38} & 26.50 &\textbf{41.70} & \textbf{35.31} & \textbf{36.42} & \textbf{35.23} & \textbf{15.24} & \textbf{43.01}&	\textbf{32.74}&	\textbf{13.03 } \\ \bottomrule
\end{tabular}
\vspace{-2.5em}
\end{table*}
}
\noindent\textbf{Stage 2: Max-Reward-Driven Reinforcement Learning.}
After the first-stage training, our model is able to generate various translation results. 
In order to directly optimize sentence-level evaluation metrics with all references, we implement max-reward reinforcement learning~\cite{RL1992, One_Ref_Not_Enough} with a policy gradient \( p_{\theta} \). 
Here, the model acts as an agent and interacts with its environment (sign videos and references). 
At each time step, the agent translates a word (action), and the generation of the end-of-sequence token results in a reward \( r \) to the agent. 
Our training objective is to minimize the negative expected reward function:
\vspace{-0.5em}
\begin{equation}
L_{\text{RL}}(\theta) = -\mathbb{E}_{\widehat{\mathbb{Y}} \sim p_{\theta}} [r(\widehat{\mathbb{Y}})]\text{.}
\vspace{-0.5em}
\end{equation}
The gradients of this non-differentiable, reward-based loss function are written by:
\vspace{-0.5em}
\begin{equation}
\nabla_{\theta} L_{\text{RL}}(\theta) = -\mathbb{E}_{\widehat{\mathbb{Y}} \sim p_{\theta}} [r(\widehat{\mathbb{Y}}) \cdot \nabla_{\theta} \log p_{\theta}(\widehat{\mathbb{Y}})]\text{.}
\vspace{-0.5em}
\end{equation}
Note that in each batch, the reward is applied to each translated sentence individually rather than the whole batch.
To enforce consistency between model performance and evaluation metrics, we utilize BLEU as the reward function.
\subsection{Training Protocols and Inference}
\textbf{Training Protocol.} In the first stage, we select one of the multi-reference training strategies to train the DivSLT model using $L_{\text{XE}}$. 
In the second stage, we fine-tune the model from the first stage using $L_{\text{RL}}$.
The model with the best performance (BLEU-MR) on the validation set is selected as our final model.

\noindent\textbf{Inference.}
In the decoding process, the beam search algorithm~\cite{NSLT,mixed_RL_3} is used. 
However, the main issue with beam search is that the diversity among the $K$ output sentences is minimal, failing to reflect the real diversity of language. 
To remedy this, we employ diverse beam search~\cite{diverse_beam_search} during inference, which partitions the beam into several groups~($G$) and applies penalties~($P$) to the candidates, thus promoting the generation of diverse translation results for SLT models.

\section{Experiments}
\label{sec:exp}

In this section, we carry out experiments to demonstrate the effectiveness of multi-reference learning on our extended SLT datasets, CSL-Daily-Div and PHOENIX14T-Div, in terms of translation accuracy and diversity. 
Without diluting the key contribution of the multiple-references, we mainly compare the SLT models trained from scratch without any extra pre-training stage.

\subsection{Implementation Details}


In the first stage, the DivSLT network is trained by a cross-entropy loss and label smoothing of 0.2 in a mini-batch size of 32. 
For both stages, we use the SGD optimizer with a momentum of 0.9 and set the learning rate to 0.03 with the cosine annealing scheduler. 
The number of training epochs in the first and second stage is set to 200 and 30, respectively.
During inference, we employ the diverse beam search strategy to obtain the Top-3 translations, which involves a diversity penalty ($P$) of 0.5, a beam width of 5, and a group size ($G$) of 5, \textit{i.e.}, each beam is divided into 5 diverse groups.

\subsection{Evaluation Metrics}
\SX{When multiple references are available, relying solely on a fixed reference can introduce biases in evaluation, especially if there is a significant divergence in vocabulary or syntax between the reference and the predicted translation. Therefore, we propose a best-matching (BM) method to assess the DivSLT models:
\vspace{-1em}
\begin{equation}
\textit{$\Psi$-BM} = \frac{1}{k} \sum_{i=1}^{k} \max_{n = 1}^N \textit{$\Psi$}(\widehat{y_{i}}, y_{i}^{n})\text{,}
\vspace{-1em}
\end{equation}
where \( \widehat{y_{i}} \) is the predicted result, \( y_{i}^{n} \) represents all references\footnote{The ground-truth and extended references.}, and \( \textit{$\Psi$} \) is the evaluation metrics, such as BLEU~\cite{bleu}, ROUGE~\cite{rouge} and BLEURT~\cite{bleuRT}.
}

We use the following two groups of metrics to assess the accuracy and diversity of Top-1 and Top-\textit{k} generated translations, respectively.

\noindent\textbf{BLEU, ROUGE, BLEURT-BM\textbackslash MR.}
For the Top-1 translation predicted by the DivSLT model, we use BLEU~\cite{bleu}, ROUGE~\cite{rouge}, and BLEURT~\cite{bleuRT,bleuRT20} to evaluate its accuracy. 
\SX{BLEU-BM, ROUGE-BM, and BLEURT-BM calculate the metric scores between the best-matching ground-truth reference and the generated one, respectively, while BLEU-MR, ROUGE-MR and BLEURT-MR calculate the metric scores between all references and the generated one, respectively.}

\noindent\textbf{pwb,~rfb-BM,~mrfb,~rfbRT-BM,~mrfbRT.} \SX{For the Top-$k$ translations predicted by the DivSLT model, we utilize pwb, rfb-BM, and rfbRT-BM metrics, where pwb, rfb and rfbRT are explained in Section~\ref{sec:dataset}. }
Besides, mrfbRT metric assesses the average semantic similarity, while mrfb computes the BLEU score with respect to all references for each generated translation.

In summary, \textbf{BLEU-BM} and \textbf{BLEURT-BM} are suggested for gauging translation accuracy and semantic precision, while \textbf{pwd} is recommended for assessing diversity.
    
{
\setlength{\tabcolsep}{4.8pt}
\begin{table*}[ht]
\centering
\tiny
\caption{DivSLT with Different Multi-Reference Training Strategies. 
``BM'', ``MR'', ``R'', ``B'', and ``BRT'' represent best-matching ground-truth, multi-reference, ROUGE, BLEU and BLEURT, respectively.
}
\vspace{-1em}
\label{diff_multi_train}
\begin{tabular}{@{}l|ccccc|cccccc@{}} \toprule
 & \multicolumn{5}{c|}{Top-3 Predictions} & \multicolumn{6}{c}{Top-1 Prediction} \\ \midrule
{\bf Multi-Reference SLT} & rfb-BM~$\uparrow$ & mrfb~$\uparrow$ & \textbf{\textcolor{blue}{pwb}}~$\downarrow$ & rfbRT-BM~$\uparrow$ & mrfbRT~$\uparrow$ & BRT-MR~$\uparrow$ & R-MR~$\uparrow$ & B-MR~$\uparrow$ & \textbf{\textcolor{blue}{BRT-BM}}~$\uparrow$ & R-BM~$\uparrow$ & B-BM~$\uparrow$ \\ \midrule 
DivSLT [Stage 1: Sample One] & 15.98 & 19.05 & \textbf{20.18} & 58.03 & 52.57 & 56.57 & 43.51 & 21.83 & 62.31&	40.78&	18.43     \\
DivSLT [Stage 2] & 17.30 & 20.57 & 25.04 & 59.65 & 54.03 & 57.10 & 44.81 & 23.21 &62.97&	42.34&	19.43   \\ \midrule
DivSLT [Stage 1: Ours]  & 20.74 & 23.96 & 22.26 & \textbf{61.89} & 54.14 & 57.15 & 48.24 & 27.69 & 64.82&	46.02&	24.20   \\
DivSLT [Stage 2] & \textbf{22.12} & \textbf{25.23} & 26.30 & 61.34 & \textbf{54.97}  &  \textbf{58.02} & \textbf{49.82} & \textbf{28.82} & \textbf{64.95}&	\textbf{47.61}&	\textbf{25.29} \\ \bottomrule
\end{tabular}
\vspace{-2em}
\end{table*}
}
\subsection{Main Results}
In Table~\ref{DSLT_DE} and Table~\ref{DSLT_ZH}, we first compare the \SX{open-source} gloss-free methods without extra pre-training stages.
GFSLT~\cite{GF_VLP} learns the association between sign language video and spoken language in an end-to-end fashion, and exhibits promising translation performance. 
Hence, our DivSLT is built on GFSLT to investigate diverse sign translation. 
In GFSLT, pre-training on the single-reference corpus can improve performance. If we pre-train our model on our multiple-reference data, this might lead to unfair comparison. 
Thus, we do not adopt the pre-training version of GFSLT.

The results show that our DivSLT baseline model, trained with multiple references, not only surpasses one-to-one SLT models in terms of diversity but also improves translation accuracy.
\SX{Specifically, benefiting from RL, the DivSLT model improves the BLEU-BM and BLEURT-BM score by approximately +5.5 (19.76$\rightarrow$25.29) and +10 (54.70$\rightarrow$64.95), respectively, on the PHOENIX14T dataset.
As for the results of the CSL-Daily dataset, the BLEU-BM and BLEURT-BM score are enhanced by about +4 (8.94$\rightarrow$13.03) and +8 (35.06$\rightarrow$43.01), respectively.}
Most importantly, the results of the generated Top-3 translations of the DivSLT surpass those of the GFSLT across the majority of diversity and accuracy evaluation metrics.
The performance of the validation sets of the two datasets is reported in the Appendix.



{
\setlength{\tabcolsep}{4.8pt}
\begin{table}[t]
\centering
\tiny
\caption{Impacts of the weights of $L_{\text{XE}}$ and $L_{\text{RL}}$ at two DivSLT training stages. ``Sep.'' represents separate.}
\vspace{-1em}
\label{loss weight}
\begin{tabular}{c|cc|c|ccc}
\toprule
&\multicolumn{2}{c|}{Loss Weight} & & \multicolumn{3}{c}{Evaluation Metric (Top-3)} \\
 & $1-\beta (L_{\text{XE}})$     &    $\beta (L_{\text{RL}})$   & Stage   &  rfb-BM~$\uparrow$ & mrfb~$\uparrow$ & \textbf{\textcolor{blue}{pwb}}~$\downarrow$   \\ \midrule 
\multirow{2}{*}{\rotatebox{90}{Sep.}} 
& 1.0              & 0.0            &  S1   &     20.74 & 23.96 & \textbf{22.26}        \\ 
& \textbf{0.0}              & \textbf{1.0}            &  \textbf{S2}  &     \textbf{22.12} & \textbf{25.23} & 26.30  \\ \midrule
\multirow{5}{*}{\rotatebox{90}{Joint}} & 0.5             & 0.5            &   S1  &     19.18   &  22.23    &    27.70                
\\ \cmidrule{2-7}
& 0.5              & 0.5            &   S2  &    21.37   &  24.66    &      23.59               \\
& 0.3              & 0.7            &   S2  &    21.40   &   24.88   &     24.37         \\
& 0.1              & 0.9            &    S2  &    21.81  &    25.10   &     25.50        \\
\bottomrule
\end{tabular}
\end{table}
}
\subsection{Ablation Study}
\noindent\textbf{Multi-reference training strategies.} 
We additionally experiment with the \textit{Sample One}~\cite{multi_reference_train_method} approach for training DivSLT.
It entails selecting a different reference for each epoch. 
Specifically, for every sign video clip, we randomly select one reference from $K$ available options in each epoch, ensuring this random choice is made anew in every subsequent epoch. The total iteration number is equal to ours.
The results in Table~\ref{diff_multi_train} illustrate that both multi-reference training strategies can enhance the diversity of translation results in the first stage,
but the DivSLT model with our training strategy produces results of greater diversity and higher accuracy.
The shuffle operation enables equal opportunities across all references in each epoch, thereby enhancing the model's ability.

\noindent\textbf{Joint training vs. Two-stage paradigm.} 
Following the approach proposed by~\citet{video_caption_RL_copy}, we attempt to jointly train the model in each stage on PHOENIX14T-Div with the mixed loss $(1-\beta) L_{\text{XE}} + \beta L_{\text{RL}}$. 
Note that Stage 2 (S2) is conducted after the Stage 1 (S1) is completed.
As shown in Table~\ref{loss weight}, the mixed loss has no effect in S1. 
We speculate that importing reward constraints before the DivSLT model converges may hurt the normal training process, as the model is not stable at the early phase.
In comparison, using mixed loss in S2 does not improve the accuracy of generated results.
Therefore, after DivSLT converges in S1, we use $L_{\text{RL}}$ exclusively to improve the accuracy of diverse translation results.
{
\setlength{\tabcolsep}{4.8pt}
\begin{table}[t]
\centering
\tiny
\caption{Impacts of different inference methods on diversity of translation results. Top-3 results are used for evaluation. ``BS'' and ``DBS'' indicate beam search and diverse beam search, respectively.
$^*$ represents our reproduced results.}
\vspace{-1em}
\label{dbs&bs}
\begin{tabular}{l|c|ccccc} \toprule
 & Infer. & rfb-BM~$\uparrow$ & mrfb~$\uparrow$ & \textbf{\textcolor{blue}{pwb}}~$\downarrow$ & rfbRT-BM~$\uparrow$ & mrfbRT~$\uparrow$  \\ \midrule
\multirow{2}{*}{GFSLT$^*$} & BS & 18.91&	20.32&	61.32&  52.31&	44.78 \\
 & DBS & 19.03 & 20.41 & 40.86 & 51.73 & 43.82  \\ \midrule
\multirow{2}{*}{DivSLT} & BS &  \textbf{23.35}&	 \textbf{26.84}&	57.14& \textbf{63.34}&	\textbf{56.12} \\
& DBS& 22.12 & 25.23 & \textbf{26.30} & 61.34 & 54.97 \\ \bottomrule
\end{tabular}
\end{table} 
}

\noindent\textbf{Different inference strategies.} 
We compare the diversity of results from both the SLT model and the DivSLT model with diverse beam search~\cite{diverse_beam_search} and beam search~\cite{mixed_RL_3} on PHOENIX14T-Div. 
Table~\ref{dbs&bs} implies that regardless of the choice of decoding strategies, DivSLT achieves better diversity than GFSLT. Diverse beam search enhances the diversity of translation results, but the most significant gain in diversity comes from our two-stage training paradigm.


\noindent\textbf{Different model architectures.}
To substantiate the efficacy of our proposed two-stage DivSLT paradigm, we integrate the paradigm with several well-performed gloss-free SLT model architectures, including RNN~\cite{NSLT}, Transformer~\cite{sltt}, \XS{TSPNet~\cite{tspnet}}, and GASLT~\cite{gloss_free_m2}.
Table~\ref{different_model} indicates that the proposed two-stage DivSLT paradigm can further improve the translation performance with most models on PHOENIX14T-Div, which verifies the effectiveness of using multi-reference training and reinforcement learning.
{
\setlength{\tabcolsep}{3.2pt}
\begin{table}[t]
\tiny
\centering
\caption{\XS{Results of various models employing the two-stage DivSLT paradigm.}}
\vspace{-1em}
\label{different_model}
\begin{tabular}{l|cccccc}
\toprule
\multicolumn{1}{c|}{} & B-BM~$\uparrow$ & B-MR~$\uparrow$ & \textbf{\textcolor{blue}{BRT-BM}}~$\uparrow$ & BRT-MR~$\uparrow$ \\ \midrule
RNN [Single Ref]  & 12.77 & 11.89 & 44.55 & 35.72 \\
RNN [Multi Ref] & \textbf{13.97} & \textbf{15.65} & \textbf{51.32} & \textbf{44.34} \\ \midrule
Transformer [Single Ref] & 10.49 & 12.14 & 41.47 &  37.34 \\
Transformer [Multi Ref] & \textbf{12.31} & \textbf{15.43} & \textbf{49.76} & \textbf{43.29} \\ \midrule
\XS{TSPNet [Single Ref]}  & \XS{14.19} & \XS{18.61} & \XS{47.73} & \XS{40.15} \\
\XS{TSPNet [Multi Ref]} &  \textbf{\XS{15.65}} & \textbf{\XS{19.97}} & \textbf{\XS{52.76}} & \textbf{\XS{49.21}} \\ \midrule
GASLT [Single Ref] & 16.45 & 18.66 & 49.07 & 41.57 \\
GASLT [Multi Ref] &  \textbf{17.53} & \textbf{20.71} & \textbf{55.37} & \textbf{51.98} \\ \midrule
GFSLT [Single Ref]  & 19.76 & 21.29 & 54.70 & 46.45  \\
GFSLT [Multi Ref] &  \textbf{25.29} & \textbf{28.82} & \textbf{64.95} & \textbf{58.02} \\ 
\bottomrule
\end{tabular}
\end{table}
}
\begin{figure}[t]
  \centering
  \includegraphics[width=1\linewidth]{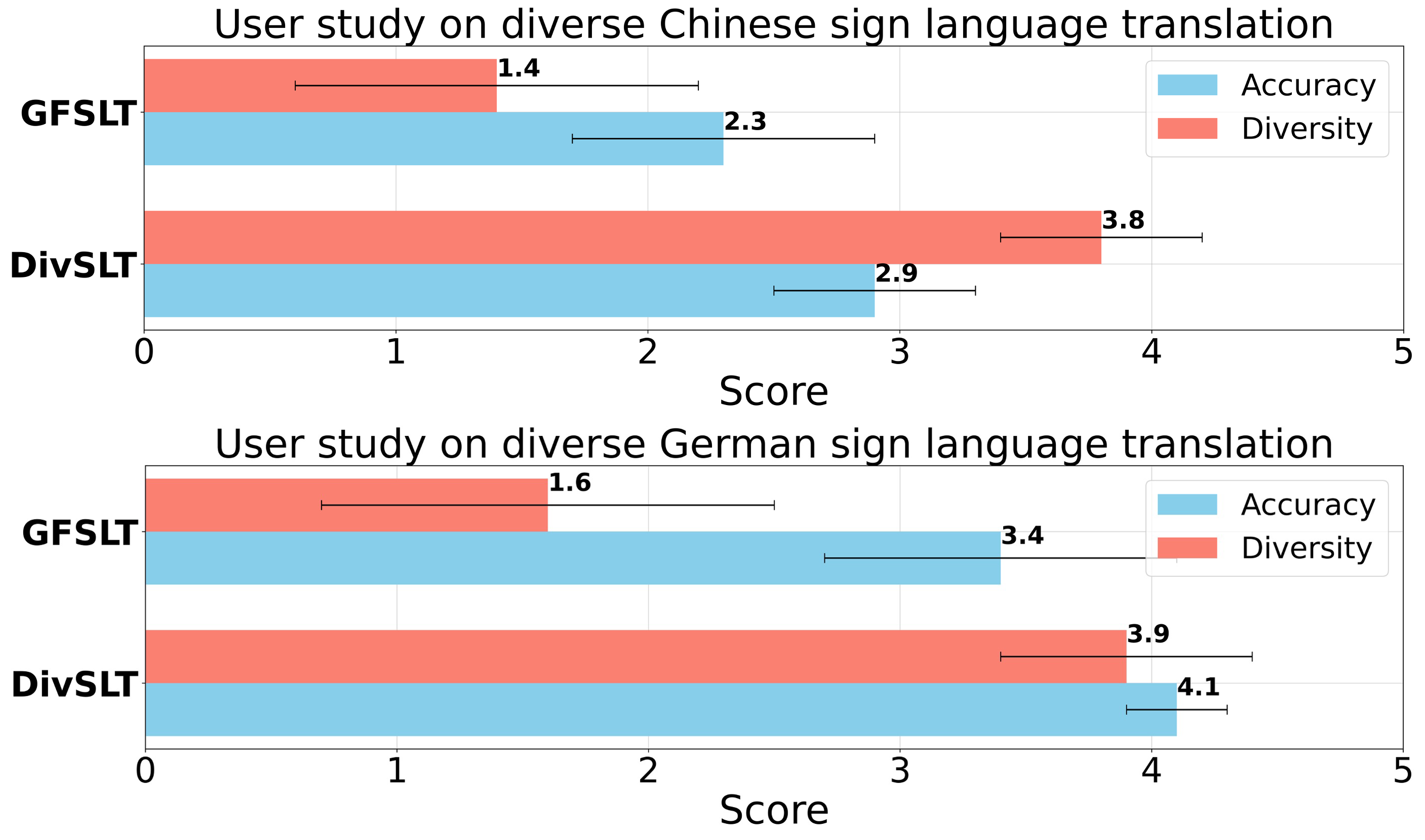}
  \vspace{-2em}
  \caption{\XS{Human evaluation on diverse sign language translation results. }
  Score 5 denotes that the translation results are completely faithful and exhibit diversity in expressions, and Score 0 indicates a complete lack of faithfulness or absence of diversity.
  }
  \label{score}
\end{figure}
\subsection{\XS{User Study}}
We randomly select \XS{100} sign video clips from both PHOENIX14T-Div and CSL-Daily-Div to compare the performance of GFSLT~\cite{GF_VLP} and DivSLT. 
Then, we invite \XS{30} native speakers to judge the translation accuracy and diversity of the generated translation results wrt. the ground-truth reference. 
For accuracy, the scores 0 to 5 represent \textit{Inaccurate}, \textit{Poor}, \textit{Fair}, \textit{Good}, \textit{Very Good}, and \textit{Excellent}, respectively.
For diversity, the scores 0 to 5 correspond to \textit{None}, \textit{Low}, \textit{Moderate}, \textit{Diverse}, \textit{Highly Diverse}, and \textit{Exceptional}, respectively.
The average value of \XS{30} evaluators over \XS{100} cases are used as the final score.
As shown in Figure~\ref{score}, it shows that DivSLT yields better translation performance for both DGS and CSL.

\subsection{Case Study}
Table~\ref{case_study} lists two cases from the extended datasets. 
In the example from PHOENIX14T-Div, ``sun and clouds alternate'' is translated as ``sometimes sun, sometimes clouds'' or ``it is changeable, sometimes the sun shines, sometimes there are clouds in the sky'' with different sentence structures but conveying the same meaning.
Diverse and accurate translation results can make it easier for people to understand sign language.
Overall, our proposed DivSLT model not only generates translation results highly similar to the ground truth but also produces translations with different expressions.
More case studies are shown in the Appendix.

{
\setlength{\tabcolsep}{0pt}
\begin{table}[t]
\centering
\tiny
\caption{Case Study. We show the differences between sentences. \textcolor{myred}{Red}: agreement with the references. \textcolor{myblue}{Blue}: correct but different words.  \textcolor{mygray}{Gray}: missing contents. }
\vspace{-1em}
\label{case_study}
\begin{tabular}{p{\linewidth}} \toprule
\textbf{Ground Truth (PHOENIX14T):} am tag wechseln sonne und wolken einander ab teilweise ist es auch längere zeit sonnig \\
(EN: The sun and clouds alternate during the day, with periods of sustained sunshine.) \\  \midrule
\textbf{Top-3 Predictions:}\\
1. \textcolor{myred}{am tag} \textcolor{myblue}{mal} \textcolor{myred}{sonne} \textcolor{myblue}{mal} \textcolor{myred}{wolken teilweise ist es auch für längere zeit sonnig}\\
(EN: \textcolor{myred}{During the day} \textcolor{myblue}{sometimes} \textcolor{myred}{sun} \textcolor{myblue}{sometimes} \textcolor{myred}{clouds sometimes it is sunny for a long time}.) \\
2. \textcolor{myred}{am tag ist es} \textcolor{myblue}{wechselhaft mal scheint} \textcolor{myred}{die sonne} \textcolor{myred}{mal sind wolken am himmel} jedoch \textcolor{myred}{auch für längere zeit die sonne}\\
(EN: \textcolor{myred}{During the day it is} \textcolor{myblue}{changeable, sometimes the} \textcolor{myred}{sun shines}, \textcolor{myblue}{sometimes there are clouds in the sky} but also \textcolor{myred}{the sun for longer periods of time}.) \\
3. \textcolor{myred}{am tag} \textcolor{myblue}{mal} \textcolor{myred}{sonne} \textcolor{myblue}{mal} \textcolor{myred}{wolken} mancherorts scheint \textcolor{myred}{die sonne auch für längere zeit}\\
(EN: \textcolor{myred}{During the day}, \textcolor{myblue}{sometimes} \textcolor{myred}{sun} \textcolor{myblue}{sometimes} \textcolor{myred}{clouds}. In some places \textcolor{myred}{the sun shines for longer periods of time}.) \\ \midrule \midrule

\textbf{Ground Truth (CSL-Daily):} \begin{CJK}{UTF8}{gbsn}天 气 预 报 明 天 下 雪 多 穿 衣 服\end{CJK} \\
(EN: The weather forecast for tomorrow is for snow, dress warmly.) \\ \midrule
\textbf{Top-3 Predictions:} \\  
\begin{CJK}{UTF8}{gbsn}1. \textcolor{myred}{天 气 预 报 明 天 有 雪 多 穿 衣 服}\end{CJK} \\
(EN: \textcolor{myred}{The weather forecast for tomorrow includes snow, dress warmly}.) \\
\begin{CJK}{UTF8}{gbsn}2. \textcolor{myred}{天 气 预 报} \textcolor{myblue}{说} \textcolor{myred}{明 天} 很 冷 要 \textcolor{myred}{多 穿} \textcolor{myblue}{几 件} \textcolor{myred}{衣 服}\end{CJK} \\
(EN: \textcolor{myred}{The weather forecast} \textcolor{myblue}{indicates} it will be very cold \textcolor{myred}{tomorrow}, \textcolor{myred}{wear several more} \textcolor{myblue}{pieces of} \textcolor{myred}{clothing}.) \\
\begin{CJK}{UTF8}{gbsn}3. \textcolor{myred}{明 天 的 天 气 预 报 显 示 会 有 雪} \textcolor{mygray}{多 穿 衣 服} \end{CJK} \\
(EN: \textcolor{myred}{Tomorrow's weather forecast shows there will be snow}, \textcolor{mygray}{dress warmly}.) \\ \bottomrule
\end{tabular}
\end{table}
}
\section{Conclusion}
\label{sec:conclusion}
In this work, we propose a new task, diverse sign language translation (DivSLT), which aims to generate diverse yet accurate translations for sign videos.
To facilitate research on this task, we first construct CSL-Daily-Div and PHOENIX14T-Div by providing multiple references to the single-reference SLT datasets. 
Then, we propose a two-stage DivSLT paradigm, the multi-reference training strategies enable an SLT model to produce diverse translations. 
Thanks to the max-reward-driven reinforcement learning, our model achieves a better trade-off between translation diversity and accuracy.
Extensive experiments on our newly extended datasets demonstrate that our method achieves better translation performance in terms of diversity and accuracy.
Our datasets and results can serve as a benchmark for DivSLT and shed some light on SLT development.

\clearpage
\section{Limitations and Future Work}
Although Diverse Sign Language Translation can yield a variety of translation outcomes, we acknowledge that our work has the following limitations, which merit further research:

\textbf{The dataset size is relatively small.}
We employ the PHOENIX14T and CSL-Daily datasets, which are widely-used in SLT research, but compared to spoken language corpora, their sizes and scopes are still limited. 
\XS{In our future work, we aim to effectively explore how to employ LLMs to enrich larger SLT datasets, such us OpenASL~\cite{openASL}, BOBSL~\cite{BOBSL} and DGS Corpus~\cite{DGS_Corpus}, with minimal human efforts, and test the DivSLT models on them.}

\textbf{Better evaluation metrics of accuracy, diversity and semantic precision need to be explored.} 
While we introduce several metrics for evaluating translation accuracy, diversity and semantic precision, a unified metric reflecting them is missing. 
We plan to develop such a metric, potentially useful for reinforcement learning to enhance translation quality.
\XS{Specifically, we will collaborate with sign language experts to provide a reasonable way to evaluate the translation. For instance, sign language experts can assist us in defining better evaluation criteria for diverse sign language translation, such as polysemy and synonym cases. With these criteria and domain knowledge provided by sign experts as prompts, we can employ LLMs to assess the results generated by a DivSLT model.}

\section*{Acknowledgement}
This research is funded in part by ARC-Discovery grant (DP220100800 to XY), ARC-DECRA grant (DE230100477 to XY) and Google Research Scholar Program.

\bibliography{custom}

\begin{thebibliography}{74}
\expandafter\ifx\csname natexlab\endcsname\relax\def\natexlab#1{#1}\fi

\bibitem[{Albanie et~al.(2021)Albanie, Varol, Momeni, Bull, Afouras, Chowdhury, Fox, Woll, Cooper, McParland, and Zisserman}]{BOBSL}
Samuel Albanie, G{\"{u}}l Varol, Liliane Momeni, Hannah Bull, Triantafyllos Afouras, Himel Chowdhury, Neil Fox, Bencie Woll, Rob Cooper, Andrew McParland, and Andrew Zisserman. 2021.
\newblock \href {http://arxiv.org/abs/2111.03635} {Bbc-oxford british sign language dataset}.
\newblock \emph{CoRR}, abs/2111.03635.

\bibitem[{Bahdanau et~al.(2017)Bahdanau, Brakel, Xu, Goyal, Lowe, Pineau, Courville, and Bengio}]{RLNMT_5}
Dzmitry Bahdanau, Philemon Brakel, Kelvin Xu, Anirudh Goyal, Ryan Lowe, Joelle Pineau, Aaron~C. Courville, and Yoshua Bengio. 2017.
\newblock \href {https://openreview.net/forum?id=SJDaqqveg} {An actor-critic algorithm for sequence prediction}.
\newblock In \emph{5th International Conference on Learning Representations, {ICLR} 2017, Toulon, France, April 24-26, 2017, Conference Track Proceedings}. OpenReview.net.

\bibitem[{Bull et~al.(2021)Bull, Afouras, Varol, Albanie, Momeni, and Zisserman}]{sign_align}
Hannah Bull, Triantafyllos Afouras, G{\"{u}}l Varol, Samuel Albanie, Liliane Momeni, and Andrew Zisserman. 2021.
\newblock \href {https://doi.org/10.1109/ICCV48922.2021.01135} {Aligning subtitles in sign language videos}.
\newblock In \emph{2021 {IEEE/CVF} International Conference on Computer Vision, {ICCV} 2021, Montreal, QC, Canada, October 10-17, 2021}, pages 11532--11541. {IEEE}.

\bibitem[{Camg{\"{o}}z et~al.(2018)Camg{\"{o}}z, Hadfield, Koller, Ney, and Bowden}]{NSLT}
Necati~Cihan Camg{\"{o}}z, Simon Hadfield, Oscar Koller, Hermann Ney, and Richard Bowden. 2018.
\newblock \href {https://doi.org/10.1109/CVPR.2018.00812} {Neural sign language translation}.
\newblock In \emph{2018 {IEEE} Conference on Computer Vision and Pattern Recognition, {CVPR} 2018, Salt Lake City, UT, USA, June 18-22, 2018}, pages 7784--7793. Computer Vision Foundation / {IEEE} Computer Society.

\bibitem[{Camg{\"{o}}z et~al.(2020)Camg{\"{o}}z, Koller, Hadfield, and Bowden}]{sltt}
Necati~Cihan Camg{\"{o}}z, Oscar Koller, Simon Hadfield, and Richard Bowden. 2020.
\newblock \href {https://doi.org/10.1109/CVPR42600.2020.01004} {Sign language transformers: Joint end-to-end sign language recognition and translation}.
\newblock In \emph{2020 {IEEE/CVF} Conference on Computer Vision and Pattern Recognition, {CVPR} 2020, Seattle, WA, USA, June 13-19, 2020}, pages 10020--10030. Computer Vision Foundation / {IEEE}.

\bibitem[{Chen et~al.(2022{\natexlab{a}})Chen, Wei, Sun, Wu, and Lin}]{mmtlb}
Yutong Chen, Fangyun Wei, Xiao Sun, Zhirong Wu, and Stephen Lin. 2022{\natexlab{a}}.
\newblock \href {https://doi.org/10.1109/CVPR52688.2022.00506} {A simple multi-modality transfer learning baseline for sign language translation}.
\newblock In \emph{{IEEE/CVF} Conference on Computer Vision and Pattern Recognition, {CVPR} 2022, New Orleans, LA, USA, June 18-24, 2022}, pages 5110--5120. {IEEE}.

\bibitem[{Chen et~al.(2022{\natexlab{b}})Chen, Zuo, Wei, Wu, Liu, and Mak}]{two_s}
Yutong Chen, Ronglai Zuo, Fangyun Wei, Yu~Wu, Shujie Liu, and Brian Mak. 2022{\natexlab{b}}.
\newblock \href {http://papers.nips.cc/paper\_files/paper/2022/hash/6cd3ac24cdb789beeaa9f7145670fcae-Abstract-Conference.html} {Two-stream network for sign language recognition and translation}.
\newblock In \emph{NeurIPS}.

\bibitem[{Deng et~al.(2009)Deng, Dong, Socher, Li, Li, and Fei{-}Fei}]{imagenet}
Jia Deng, Wei Dong, Richard Socher, Li{-}Jia Li, Kai Li, and Li~Fei{-}Fei. 2009.
\newblock \href {https://doi.org/10.1109/CVPR.2009.5206848} {Imagenet: {A} large-scale hierarchical image database}.
\newblock In \emph{2009 {IEEE} Computer Society Conference on Computer Vision and Pattern Recognition {(CVPR} 2009), 20-25 June 2009, Miami, Florida, {USA}}, pages 248--255. {IEEE} Computer Society.

\bibitem[{Devlin et~al.(2019)Devlin, Chang, Lee, and Toutanova}]{bert}
Jacob Devlin, Ming{-}Wei Chang, Kenton Lee, and Kristina Toutanova. 2019.
\newblock {BERT:} pre-training of deep bidirectional transformers for language understanding.
\newblock In \emph{Proceedings of the 2019 Conference of the North American Chapter of the Association for Computational Linguistics}, pages 4171--4186. Association for Computational Linguistics.

\bibitem[{Dreyer and Marcu(2012)}]{variability&expressiveness}
Markus Dreyer and Daniel Marcu. 2012.
\newblock \href {https://aclanthology.org/N12-1017/} {Hyter: Meaning-equivalent semantics for translation evaluation}.
\newblock In \emph{Human Language Technologies: Conference of the North American Chapter of the Association of Computational Linguistics, Proceedings, June 3-8, 2012, Montr{\'{e}}al, Canada}, pages 162--171. The Association for Computational Linguistics.

\bibitem[{Duarte et~al.(2021)Duarte, Palaskar, Ventura, Ghadiyaram, DeHaan, Metze, Torres, and Gir{\'{o}}{-}i{-}Nieto}]{how2sign}
Amanda~Cardoso Duarte, Shruti Palaskar, Lucas Ventura, Deepti Ghadiyaram, Kenneth DeHaan, Florian Metze, Jordi Torres, and Xavier Gir{\'{o}}{-}i{-}Nieto. 2021.
\newblock \href {https://doi.org/10.1109/CVPR46437.2021.00276} {How2sign: {A} large-scale multimodal dataset for continuous american sign language}.
\newblock In \emph{{IEEE} Conference on Computer Vision and Pattern Recognition, {CVPR} 2021, virtual, June 19-25, 2021}, pages 2735--2744. Computer Vision Foundation / {IEEE}.

\bibitem[{Edunov et~al.(2018)Edunov, Ott, Auli, Grangier, and Ranzato}]{RLNMT_3}
Sergey Edunov, Myle Ott, Michael Auli, David Grangier, and Marc'Aurelio Ranzato. 2018.
\newblock \href {https://doi.org/10.18653/V1/N18-1033} {Classical structured prediction losses for sequence to sequence learning}.
\newblock In \emph{Proceedings of the 2018 Conference of the North American Chapter of the Association for Computational Linguistics: Human Language Technologies, {NAACL-HLT} 2018, New Orleans, Louisiana, USA, June 1-6, 2018, Volume 1 (Long Papers)}, pages 355--364. Association for Computational Linguistics.

\bibitem[{Hanke et~al.(2020)Hanke, Schulder, Konrad, and Jahn}]{DGS_Corpus}
Thomas Hanke, Marc Schulder, Reiner Konrad, and Elena Jahn. 2020.
\newblock Extending the public dgs corpus in size and depth.
\newblock In \emph{sign-lang@ LREC 2020}, pages 75--82. European Language Resources Association (ELRA).

\bibitem[{He et~al.(2016{\natexlab{a}})He, Xia, Qin, Wang, Yu, Liu, and Ma}]{RLNMT_4}
Di~He, Yingce Xia, Tao Qin, Liwei Wang, Nenghai Yu, Tie{-}Yan Liu, and Wei{-}Ying Ma. 2016{\natexlab{a}}.
\newblock \href {https://proceedings.neurips.cc/paper/2016/hash/5b69b9cb83065d403869739ae7f0995e-Abstract.html} {Dual learning for machine translation}.
\newblock In \emph{Advances in Neural Information Processing Systems 29: Annual Conference on Neural Information Processing Systems 2016, December 5-10, 2016, Barcelona, Spain}, pages 820--828.

\bibitem[{He et~al.(2016{\natexlab{b}})He, Zhang, Ren, and Sun}]{resnet}
Kaiming He, Xiangyu Zhang, Shaoqing Ren, and Jian Sun. 2016{\natexlab{b}}.
\newblock \href {https://doi.org/10.1109/CVPR.2016.90} {Deep residual learning for image recognition}.
\newblock In \emph{2016 {IEEE} Conference on Computer Vision and Pattern Recognition, {CVPR} 2016, Las Vegas, NV, USA, June 27-30, 2016}, pages 770--778. {IEEE} Computer Society.

\bibitem[{Hu et~al.(2023)Hu, Pu, Zhou, Fang, and Li}]{cslr3}
Hezhen Hu, Junfu Pu, Wengang Zhou, Hang Fang, and Houqiang Li. 2023.
\newblock \href {https://doi.org/10.1109/TMM.2023.3268368} {Prior-aware cross modality augmentation learning for continuous sign language recognition}.
\newblock \emph{IEEE Transactions on Multimedia}, pages 1--14.

\bibitem[{Jiao et~al.(2023)Jiao, Wang, Huang, Wang, and Tu}]{chatgpt_good_translator}
Wenxiang Jiao, Wenxuan Wang, Jen-tse Huang, Xing Wang, and Zhaopeng Tu. 2023.
\newblock Is chatgpt a good translator? a preliminary study.
\newblock \emph{arXiv preprint arXiv:2301.08745}.

\bibitem[{Karpathy and Fei-Fei(2015)}]{image_caption_sample_one}
Andrej Karpathy and Li~Fei-Fei. 2015.
\newblock Deep visual-semantic alignments for generating image descriptions.
\newblock In \emph{Proceedings of the IEEE conference on computer vision and pattern recognition}, pages 3128--3137.

\bibitem[{Khayrallah et~al.(2020)Khayrallah, Thompson, Post, and Koehn}]{multi_ref_improve_NMT}
Huda Khayrallah, Brian Thompson, Matt Post, and Philipp Koehn. 2020.
\newblock \href {https://doi.org/10.18653/V1/2020.EMNLP-MAIN.7} {Simulated multiple reference training improves low-resource machine translation}.
\newblock In \emph{Proceedings of the 2020 Conference on Empirical Methods in Natural Language Processing, {EMNLP} 2020, Online, November 16-20, 2020}, pages 82--89. Association for Computational Linguistics.

\bibitem[{Ko et~al.(2018)Ko, Kim, Jung, and Cho}]{KSL}
Sang{-}Ki Ko, Chang~Jo Kim, Hyedong Jung, and Choong~Sang Cho. 2018.
\newblock \href {http://arxiv.org/abs/1811.11436} {Neural sign language translation based on human keypoint estimation}.
\newblock \emph{CoRR}, abs/1811.11436.

\bibitem[{Lachaux et~al.(2020)Lachaux, Joulin, and Lample}]{LachauxJL2020}
Marie{-}Anne Lachaux, Armand Joulin, and Guillaume Lample. 2020.
\newblock \href {https://doi.org/10.18653/V1/2020.FINDINGS-EMNLP.256} {Target conditioning for one-to-many generation}.
\newblock In \emph{Findings of the Association for Computational Linguistics: {EMNLP} 2020, Online Event, 16-20 November 2020}, volume {EMNLP} 2020 of \emph{Findings of {ACL}}, pages 2853--2862. Association for Computational Linguistics.

\bibitem[{Li et~al.(2020{\natexlab{a}})Li, Rodriguez, Yu, and Li}]{wlasl}
Dongxu Li, Cristian Rodriguez, Xin Yu, and Hongdong Li. 2020{\natexlab{a}}.
\newblock Word-level deep sign language recognition from video: A new large-scale dataset and methods comparison.
\newblock In \emph{The IEEE Winter Conference on Applications of Computer Vision}, pages 1459--1469.

\bibitem[{Li et~al.(2020{\natexlab{b}})Li, Xu, Yu, Zhang, Swift, Suominen, and Li}]{tspnet}
Dongxu Li, Chenchen Xu, Xin Yu, Kaihao Zhang, Benjamin Swift, Hanna Suominen, and Hongdong Li. 2020{\natexlab{b}}.
\newblock \href {https://proceedings.neurips.cc/paper/2020/hash/8c00dee24c9878fea090ed070b44f1ab-Abstract.html} {Tspnet: Hierarchical feature learning via temporal semantic pyramid for sign language translation}.
\newblock In \emph{Advances in Neural Information Processing Systems 33: Annual Conference on Neural Information Processing Systems 2020, NeurIPS 2020, December 6-12, 2020, virtual}.

\bibitem[{Li et~al.(2016)Li, Monroe, and Jurafsky}]{li2016simple}
Jiwei Li, Will Monroe, and Dan Jurafsky. 2016.
\newblock A simple, fast diverse decoding algorithm for neural generation.
\newblock \emph{arXiv preprint arXiv:1611.08562}.

\bibitem[{Lin(2004)}]{rouge}
Chin-Yew Lin. 2004.
\newblock Rouge: A package for automatic evaluation of summaries.
\newblock In \emph{Text summarization branches out}, pages 74--81.

\bibitem[{Lin et~al.(2022)Lin, Yang, Yao, Liu, Zhang, Xie, Zhang, and Su}]{Multi_Candidate_Opt}
Huan Lin, Baosong Yang, Liang Yao, Dayiheng Liu, Haibo Zhang, Jun Xie, Min Zhang, and Jinsong Su. 2022.
\newblock \href {https://doi.org/10.18653/V1/2022.FINDINGS-NAACL.200} {Bridging the gap between training and inference: Multi-candidate optimization for diverse neural machine translation}.
\newblock In \emph{Findings of the Association for Computational Linguistics: {NAACL} 2022, Seattle, WA, United States, July 10-15, 2022}, pages 2622--2632. Association for Computational Linguistics.

\bibitem[{Lin et~al.()Lin, Yao, Yang, Liu, Zhang, Luo, Huang, and Su}]{DivNMT_4}
Huan Lin, Liang Yao, Baosong Yang, Dayiheng Liu, Haibo Zhang, Weihua Luo, Degen Huang, and Jinsong Su.
\newblock Towards user-driven neural machine translation.
\newblock In \emph{Proceedings of the 59th Annual Meeting of the Association for Computational Linguistics and the 11th International Joint Conference on Natural Language Processing, {ACL/IJCNLP} 2021}. Association for Computational Linguistics.

\bibitem[{Lin et~al.(2023)Lin, Wang, Zhu, Sun, Zhang, and Yang}]{e2e_pose_SLT}
Kezhou Lin, Xiaohan Wang, Linchao Zhu, Ke~Sun, Bang Zhang, and Yi~Yang. 2023.
\newblock Gloss-free end-to-end sign language translation.
\newblock In \emph{Proceedings of the 61st Annual Meeting of the Association for Computational Linguistics (Volume 1: Long Papers), {ACL} 2023, Toronto, Canada, July 9-14, 2023}, pages 12904--12916. Association for Computational Linguistics.

\bibitem[{Liu et~al.(2023)Liu, Han, Ma, Zhang, Yang, Tian, He, Li, He, Liu et~al.}]{llm_summary}
Yiheng Liu, Tianle Han, Siyuan Ma, Jiayue Zhang, Yuanyuan Yang, Jiaming Tian, Hao He, Antong Li, Mengshen He, Zhengliang Liu, et~al. 2023.
\newblock Summary of chatgpt-related research and perspective towards the future of large language models.
\newblock \emph{Meta-Radiology}, page 100017.

\bibitem[{Liu et~al.(2020)Liu, Gu, Goyal, Li, Edunov, Ghazvininejad, Lewis, and Zettlemoyer}]{mbart}
Yinhan Liu, Jiatao Gu, Naman Goyal, Xian Li, Sergey Edunov, Marjan Ghazvininejad, Mike Lewis, and Luke Zettlemoyer. 2020.
\newblock \href {https://doi.org/10.1162/TACL\_A\_00343} {Multilingual denoising pre-training for neural machine translation}.
\newblock \emph{Trans. Assoc. Comput. Linguistics}, 8:726--742.

\bibitem[{Mehrish et~al.(2023)Mehrish, Majumder, Bharadwaj, Mihalcea, and Poria}]{mmnlg3}
Ambuj Mehrish, Navonil Majumder, Rishabh Bharadwaj, Rada Mihalcea, and Soujanya Poria. 2023.
\newblock A review of deep learning techniques for speech processing.
\newblock \emph{Information Fusion}, page 101869.

\bibitem[{Michel and Neubig(2018{\natexlab{a}})}]{DivNMT_2}
Paul Michel and Graham Neubig. 2018{\natexlab{a}}.
\newblock \href {https://doi.org/10.18653/V1/P18-2050} {Extreme adaptation for personalized neural machine translation}.
\newblock In \emph{Proceedings of the 56th Annual Meeting of the Association for Computational Linguistics, {ACL} 2018, Melbourne, Australia, July 15-20, 2018, Volume 2: Short Papers}, pages 312--318. Association for Computational Linguistics.

\bibitem[{Michel and Neubig(2018{\natexlab{b}})}]{DivNMT_3}
Paul Michel and Graham Neubig. 2018{\natexlab{b}}.
\newblock \href {https://doi.org/10.18653/V1/P18-2050} {Extreme adaptation for personalized neural machine translation}.
\newblock In \emph{Proceedings of the 56th Annual Meeting of the Association for Computational Linguistics, {ACL} 2018, Melbourne, Australia, July 15-20, 2018, Volume 2: Short Papers}, pages 312--318. Association for Computational Linguistics.

\bibitem[{Moryossef et~al.(2021)Moryossef, Yin, Neubig, and Goldberg}]{g2t1}
Amit Moryossef, Kayo Yin, Graham Neubig, and Yoav Goldberg. 2021.
\newblock \href {https://aclanthology.org/2021.mtsummit-at4ssl.1} {Data augmentation for sign language gloss translation}.
\newblock In \emph{Proceedings of the 1st International Workshop on Automatic Translation for Signed and Spoken Languages}, pages 1--11. Association for Machine Translation in the Americas.

\bibitem[{{OpenAI}(2023)}]{openai2023chatgpt}
{OpenAI}. 2023.
\newblock Chatgpt.
\newblock Available at \url{https://www.openai.com/chatgpt/} (Accessed: 2023-10-30).

\bibitem[{Papineni et~al.(2002)Papineni, Roukos, Ward, and Zhu}]{bleu}
Kishore Papineni, Salim Roukos, Todd Ward, and Wei-Jing Zhu. 2002.
\newblock Bleu: a method for automatic evaluation of machine translation.
\newblock In \emph{Proceedings of the 40th annual meeting of the Association for Computational Linguistics}, pages 311--318.

\bibitem[{Pasunuru and Bansal(2017)}]{video_caption_RL_copy}
Ramakanth Pasunuru and Mohit Bansal. 2017.
\newblock Reinforced video captioning with entailment rewards.
\newblock In \emph{Proceedings of the 2017 Conference on Empirical Methods in Natural Language Processing, {EMNLP} 2017, Copenhagen, Denmark, September 9-11, 2017}.

\bibitem[{Paulus et~al.(2018)Paulus, Xiong, and Socher}]{mixed_RL_2}
Romain Paulus, Caiming Xiong, and Richard Socher. 2018.
\newblock A deep reinforced model for abstractive summarization.
\newblock In \emph{6th International Conference on Learning Representations, {ICLR} 2018}.

\bibitem[{Pu et~al.(2021)Pu, Chung, Parikh, Gehrmann, and Sellam}]{bleuRT20}
Amy Pu, Hyung~Won Chung, Ankur~P Parikh, Sebastian Gehrmann, and Thibault Sellam. 2021.
\newblock Learning compact metrics for mt.
\newblock In \emph{Proceedings of EMNLP}.

\bibitem[{Radford et~al.(2021)Radford, Kim, Hallacy, Ramesh, Goh, Agarwal, Sastry, Askell, Mishkin, Clark, Krueger, and Sutskever}]{clip}
Alec Radford, Jong~Wook Kim, Chris Hallacy, Aditya Ramesh, Gabriel Goh, Sandhini Agarwal, Girish Sastry, Amanda Askell, Pamela Mishkin, Jack Clark, Gretchen Krueger, and Ilya Sutskever. 2021.
\newblock Learning transferable visual models from natural language supervision.
\newblock In \emph{Proceedings of the 38th International Conference on Machine Learning, {ICML} 2021, 18-24 July 2021, Virtual Event}.

\bibitem[{Ranzato et~al.(2016)Ranzato, Chopra, Auli, and Zaremba}]{RLNMT_1}
Marc'Aurelio Ranzato, Sumit Chopra, Michael Auli, and Wojciech Zaremba. 2016.
\newblock \href {http://arxiv.org/abs/1511.06732} {Sequence level training with recurrent neural networks}.
\newblock In \emph{4th International Conference on Learning Representations, {ICLR} 2016, San Juan, Puerto Rico, May 2-4, 2016, Conference Track Proceedings}.

\bibitem[{Sellam et~al.(2020)Sellam, Das, and Parikh}]{bleuRT}
Thibault Sellam, Dipanjan Das, and Ankur~P. Parikh. 2020.
\newblock \href {https://doi.org/10.18653/V1/2020.ACL-MAIN.704} {{BLEURT:} learning robust metrics for text generation}.
\newblock In \emph{Proceedings of the 58th Annual Meeting of the Association for Computational Linguistics, {ACL} 2020, Online, July 5-10, 2020}, pages 7881--7892. Association for Computational Linguistics.

\bibitem[{Shao et~al.(2022)Shao, Wu, and Feng}]{One_Ref_Not_Enough}
Chenze Shao, Xuanfu Wu, and Yang Feng. 2022.
\newblock One reference is not enough: Diverse distillation with reference selection for non-autoregressive translation.
\newblock In \emph{Proceedings of the 2022 Conference of the North American Chapter of the Association for Computational Linguistics: Human Language Technologies, {NAACL}}. Association for Computational Linguistics.

\bibitem[{Shen et~al.(2021)Shen, Zhan, Shen, Song, and Zhao}]{nlg_1}
Lei Shen, Haolan Zhan, Xin Shen, Yonghao Song, and Xiaofang Zhao. 2021.
\newblock \href {https://doi.org/10.1145/3474085.3475568} {Text is {NOT} enough: Integrating visual impressions into open-domain dialogue generation}.
\newblock In \emph{{MM} '21: {ACM} Multimedia Conference, Virtual Event, China, October 20 - 24, 2021}, pages 4287--4296. {ACM}.

\bibitem[{Shen et~al.(2019)Shen, Ott, Auli, and Ranzato}]{shen2019}
Tianxiao Shen, Myle Ott, Michael Auli, and Marc'Aurelio Ranzato. 2019.
\newblock Mixture models for diverse machine translation: Tricks of the trade.
\newblock In \emph{Proceedings of the 36th International Conference on Machine Learning, {ICML}}, Proceedings of Machine Learning Research.

\bibitem[{Shen et~al.(2023)Shen, Yuan, Sheng, Du, and Yu}]{shen2023auslan}
Xin Shen, Shaozu Yuan, Hongwei Sheng, Heming Du, and Xin Yu. 2023.
\newblock Auslan-daily: Australian sign language translation for daily communication and news.
\newblock In \emph{Thirty-seventh Conference on Neural Information Processing Systems Datasets and Benchmarks Track}.

\bibitem[{Sheng et~al.(2024)Sheng, Shen, Du, Zhang, Huang, and Yu}]{sheng2024ai}
Hongwei Sheng, Xin Shen, Heming Du, Hu~Zhang, Zi~Huang, and Xin Yu. 2024.
\newblock Ai empowered auslan learning for parents of deaf children and children of deaf adults.
\newblock \emph{AI and Ethics}, pages 1--11.

\bibitem[{Shi et~al.(2022)Shi, Brentari, Shakhnarovich, and Livescu}]{openASL}
Bowen Shi, Diane Brentari, Greg Shakhnarovich, and Karen Livescu. 2022.
\newblock Open-domain sign language translation learned from online video.
\newblock \emph{arXiv preprint arXiv:2205.12870}.

\bibitem[{Shu et~al.(2019)Shu, Nakayama, and Cho}]{shu2019}
Raphael Shu, Hideki Nakayama, and Kyunghyun Cho. 2019.
\newblock \href {https://doi.org/10.18653/V1/P19-1177} {Generating diverse translations with sentence codes}.
\newblock In \emph{Proceedings of the 57th Conference of the Association for Computational Linguistics, {ACL} 2019, Florence, Italy, July 28- August 2, 2019, Volume 1: Long Papers}, pages 1823--1827. Association for Computational Linguistics.

\bibitem[{Su et~al.(2017)Su, Tan, Xiong, Ji, Shi, and Liu}]{DivNMT_1}
Jinsong Su, Zhixing Tan, Deyi Xiong, Rongrong Ji, Xiaodong Shi, and Yang Liu. 2017.
\newblock \href {https://doi.org/10.1609/AAAI.V31I1.10968} {Lattice-based recurrent neural network encoders for neural machine translation}.
\newblock In \emph{Proceedings of the Thirty-First {AAAI} Conference on Artificial Intelligence, February 4-9, 2017, San Francisco, California, {USA}}, pages 3302--3308. {AAAI} Press.

\bibitem[{Sun et~al.(2020)Sun, Huang, Wei, Dai, and Chen}]{sun2020}
Zewei Sun, Shujian Huang, Hao{-}Ran Wei, Xinyu Dai, and Jiajun Chen. 2020.
\newblock \href {https://doi.org/10.1609/AAAI.V34I05.6429} {Generating diverse translation by manipulating multi-head attention}.
\newblock In \emph{The Thirty-Fourth {AAAI} Conference on Artificial Intelligence, {AAAI} 2020}, pages 8976--8983. {AAAI} Press.

\bibitem[{Touvron et~al.(2023)Touvron, Martin, Stone, Albert, Almahairi, Babaei, Bashlykov, Batra, Bhargava et~al.}]{llama2}
Hugo Touvron, Louis Martin, Kevin Stone, Peter Albert, Amjad Almahairi, Yasmine Babaei, Nikolay Bashlykov, Soumya Batra, Prajjwal Bhargava, et~al. 2023.
\newblock \href {https://doi.org/10.48550/ARXIV.2307.09288} {Llama 2: Open foundation and fine-tuned chat models}.
\newblock \emph{CoRR}, abs/2307.09288.

\bibitem[{Vijayakumar et~al.(2018)Vijayakumar, Cogswell, Selvaraju, Sun, Lee, Crandall, and Batra}]{vijayakumar2018}
Ashwin~K. Vijayakumar, Michael Cogswell, Ramprasaath~R. Selvaraju, Qing Sun, Stefan Lee, David~J. Crandall, and Dhruv Batra. 2018.
\newblock \href {https://doi.org/10.1609/AAAI.V32I1.12340} {Diverse beam search for improved description of complex scenes}.
\newblock In \emph{Proceedings of the Thirty-Second {AAAI}}, pages 7371--7379. {AAAI} Press.

\bibitem[{Vijayakumar et~al.(2016)Vijayakumar, Cogswell, Selvaraju, Sun, Lee, Crandall, and Batra}]{diverse_beam_search}
Ashwin~K Vijayakumar, Michael Cogswell, Ramprasath~R Selvaraju, Qing Sun, Stefan Lee, David Crandall, and Dhruv Batra. 2016.
\newblock Diverse beam search: Decoding diverse solutions from neural sequence models.
\newblock \emph{arXiv preprint arXiv:1610.02424}.

\bibitem[{Wang et~al.(2022)Wang, Jiao, Hao, Wang, Shi, Tu, and Lyu}]{human_eva}
Wenxuan Wang, Wenxiang Jiao, Yongchang Hao, Xing Wang, Shuming Shi, Zhaopeng Tu, and Michael Lyu. 2022.
\newblock Understanding and improving sequence-to-sequence pretraining for neural machine translation.
\newblock \emph{arXiv preprint arXiv:2203.08442}.

\bibitem[{Wei et~al.(2023)Wei, Yuan, Chen, Shen, Wang, Shen, and Yan}]{vc_1}
Yiwei Wei, Shaozu Yuan, Meng Chen, Xin Shen, Longbiao Wang, Lei Shen, and Zhiling Yan. 2023.
\newblock \href {https://doi.org/10.1016/J.NEUCOM.2023.126523} {Mpp-net: Multi-perspective perception network for dense video captioning}.
\newblock \emph{Neurocomputing}, 552:126523.

\bibitem[{Williams(1992)}]{RL1992}
Ronald~J. Williams. 1992.
\newblock \href {https://doi.org/10.1007/BF00992696} {Simple statistical gradient-following algorithms for connectionist reinforcement learning}.
\newblock \emph{Mach. Learn.}, 8:229--256.

\bibitem[{Wu et~al.(2017)Wu, Zhao, Qin, Lai, and Liu}]{RLNMT_2}
Lijun Wu, Li~Zhao, Tao Qin, Jianhuang Lai, and Tie{-}Yan Liu. 2017.
\newblock \href {https://doi.org/10.24963/IJCAI.2017/432} {Sequence prediction with unlabeled data by reward function learning}.
\newblock In \emph{Proceedings of the Twenty-Sixth International Joint Conference on Artificial Intelligence, {IJCAI} 2017, Melbourne, Australia, August 19-25, 2017}, pages 3098--3104. ijcai.org.

\bibitem[{Wu et~al.(2020)Wu, Feng, and Shao}]{wu2020}
Xuanfu Wu, Yang Feng, and Chenze Shao. 2020.
\newblock Generating diverse translation from model distribution with dropout.
\newblock In \emph{Proceedings of the 2020 Conference on Empirical Methods in Natural Language Processing, {EMNLP} 2020, Online, November 16-20, 2020}. Association for Computational Linguistics.

\bibitem[{Wu et~al.(2016)Wu, Schuster, Chen, Le, Norouzi, Macherey, Krikun et~al.}]{mixed_RL_3}
Yonghui Wu, Mike Schuster, Zhifeng Chen, Quoc~V. Le, Mohammad Norouzi, Wolfgang Macherey, Maxim Krikun, et~al. 2016.
\newblock \href {http://arxiv.org/abs/1609.08144} {Google's neural machine translation system: Bridging the gap between human and machine translation}.
\newblock \emph{CoRR}, abs/1609.08144.

\bibitem[{Yang et~al.(2023)Yang, Nagrani, Seo, Miech, Pont-Tuset, Laptev, Sivic, and Schmid}]{mmnlg1}
Antoine Yang, Arsha Nagrani, Paul~Hongsuck Seo, Antoine Miech, Jordi Pont-Tuset, Ivan Laptev, Josef Sivic, and Cordelia Schmid. 2023.
\newblock Vid2seq: Large-scale pretraining of a visual language model for dense video captioning.
\newblock In \emph{Proceedings of the IEEE/CVF Conference on Computer Vision and Pattern Recognition (CVPR)}, pages 10714--10726.

\bibitem[{Yin et~al.(2023)Yin, Zhong, Tang, Jin, Jin, and Zhao}]{gloss_free_m2}
Aoxiong Yin, Tianyun Zhong, Li~Tang, Weike Jin, Tao Jin, and Zhou Zhao. 2023.
\newblock Gloss attention for gloss-free sign language translation.
\newblock In \emph{Proceedings of the IEEE/CVF Conference on Computer Vision and Pattern Recognition}, pages 2551--2562.

\bibitem[{Yin and Read(2020)}]{yin2020better}
Kayo Yin and Jesse Read. 2020.
\newblock Better sign language translation with stmc-transformer.
\newblock In \emph{Proceedings of the 28th International Conference on Computational Linguistics}, pages 5975--5989.

\bibitem[{Zeng et~al.(2022)Zeng, Liu, Du, Wang, Lai, Ding, Yang, Xu, Zheng, Xia et~al.}]{glm_130b}
Aohan Zeng, Xiao Liu, Zhengxiao Du, Zihan Wang, Hanyu Lai, Ming Ding, Zhuoyi Yang, Yifan Xu, Wendi Zheng, Xiao Xia, et~al. 2022.
\newblock Glm-130b: An open bilingual pre-trained model.
\newblock \emph{arXiv preprint arXiv:2210.02414}.

\bibitem[{Zhang et~al.(2023{\natexlab{a}})Zhang, M{\"{u}}ller, and Sennrich}]{iclr23}
Biao Zhang, Mathias M{\"{u}}ller, and Rico Sennrich. 2023{\natexlab{a}}.
\newblock \href {https://doi.org/10.48550/arXiv.2305.01778} {{SLTUNET:} {A} simple unified model for sign language translation}.
\newblock \emph{CoRR}, abs/2305.01778.

\bibitem[{Zhang et~al.(2023{\natexlab{b}})Zhang, Wu, Sun, Tu, Lu, Min, Chen, and Zhai}]{mmnlg2}
Zicheng Zhang, Wei Wu, Wei Sun, Danyang Tu, Wei Lu, Xiongkuo Min, Ying Chen, and Guangtao Zhai. 2023{\natexlab{b}}.
\newblock Md-vqa: Multi-dimensional quality assessment for ugc live videos.
\newblock In \emph{Proceedings of the IEEE/CVF Conference on Computer Vision and Pattern Recognition (CVPR)}, pages 1746--1755.

\bibitem[{Zhao et~al.(2021)Zhao, Qi, Zhou, Duan, Zhou, and Li}]{gloss_free_m1}
Jian Zhao, Weizhen Qi, Wengang Zhou, Nan Duan, Ming Zhou, and Houqiang Li. 2021.
\newblock Conditional sentence generation and cross-modal reranking for sign language translation.
\newblock \emph{IEEE Transactions on Multimedia}, 24:2662--2672.

\bibitem[{Zheng et~al.(2021)Zheng, Chen, Wu, Shi, and Kamal}]{gloss_free_m9}
Jiangbin Zheng, Yidong Chen, Chong Wu, Xiaodong Shi, and Suhail~Muhammad Kamal. 2021.
\newblock \href {https://doi.org/10.1016/j.neucom.2021.08.079} {Enhancing neural sign language translation by highlighting the facial expression information}.
\newblock \emph{Neurocomputing}, 464:462--472.

\bibitem[{Zheng et~al.(2023)Zheng, Wang, Tan, Li, Wang, Xia, Chen, and Li}]{cslr2}
Jiangbin Zheng, Yile Wang, Cheng Tan, Siyuan Li, Ge~Wang, Jun Xia, Yidong Chen, and Stan~Z Li. 2023.
\newblock Cvt-slr: Contrastive visual-textual transformation for sign language recognition with variational alignment.
\newblock In \emph{Proceedings of the IEEE/CVF Conference on Computer Vision and Pattern Recognition}, pages 23141--23150.

\bibitem[{Zheng et~al.(2018)Zheng, Ma, and Huang}]{multi_reference_train_method}
Renjie Zheng, Mingbo Ma, and Liang Huang. 2018.
\newblock \href {https://doi.org/10.18653/V1/D18-1357} {Multi-reference training with pseudo-references for neural translation and text generation}.
\newblock In \emph{Proceedings of the 2018 Conference on Empirical Methods in Natural Language Processing, Brussels, Belgium, October 31 - November 4, 2018}, pages 3188--3197. Association for Computational Linguistics.

\bibitem[{Zhou et~al.(2023)Zhou, Chen, Clap{\'e}s, Wan, Liang, Escalera, Lei, and Zhang}]{GF_VLP}
Benjia Zhou, Zhigang Chen, Albert Clap{\'e}s, Jun Wan, Yanyan Liang, Sergio Escalera, Zhen Lei, and Du~Zhang. 2023.
\newblock Gloss-free sign language translation: Improving from visual-language pretraining.
\newblock In \emph{Proceedings of the IEEE/CVF International Conference on Computer Vision}, pages 20871--20881.

\bibitem[{Zhou et~al.(2021{\natexlab{a}})Zhou, Zhou, Qi, Pu, and Li}]{CSL_Daily}
Hao Zhou, Wengang Zhou, Weizhen Qi, Junfu Pu, and Houqiang Li. 2021{\natexlab{a}}.
\newblock \href {https://doi.org/10.1109/CVPR46437.2021.00137} {Improving sign language translation with monolingual data by sign back-translation}.
\newblock In \emph{{IEEE} Conference on Computer Vision and Pattern Recognition, {CVPR} 2021, virtual, June 19-25, 2021}, pages 1316--1325. Computer Vision Foundation / {IEEE}.

\bibitem[{Zhou et~al.(2021{\natexlab{b}})Zhou, Zhou, Zhou, and Li}]{zhou2021spatial}
Hao Zhou, Wengang Zhou, Yun Zhou, and Houqiang Li. 2021{\natexlab{b}}.
\newblock Spatial-temporal multi-cue network for sign language recognition and translation.
\newblock \emph{IEEE Transactions on Multimedia}, 24:768--779.

\bibitem[{Zuo and Mak(2022)}]{cslr1}
Ronglai Zuo and Brian Mak. 2022.
\newblock \href {https://doi.org/10.1109/CVPR52688.2022.00507} {C\({}^{\mbox{2}}\)slr: Consistency-enhanced continuous sign language recognition}.
\newblock In \emph{{IEEE/CVF} Conference on Computer Vision and Pattern Recognition, {CVPR} 2022, New Orleans, LA, USA, June 18-24, 2022}, pages 5121--5130. {IEEE}.

\end{thebibliography}

\appendix

\section*{Appendix}
\setcounter{page}{1}
\label{sec:X_suppl}
\setcounter{section}{0} 
\renewcommand{\thesection}{\Alph{section}}  

\section{LLM Comparison for Multi-reference Generation Results}
\label{sec:llm Comparison}
We conduct a manual review of the generated outputs by different LLMs. 
We engage three native speakers to pinpoint errors in the translations, including instances of under-translation, over-translation, and mis-translation~\cite{human_eva}. 
Concurrently, this process evaluates the diversity and fidelity of the generated results.
Based on all the evaluation criteria, the native speakers rank the translation outputs of different LLMs, with 1 as the best SLT.
As indicated in Table~\ref{multi_ref_case}, for the CSL-Daily dataset, we employ multilingual models, for example, ChatGPT-4~\cite{openai2023chatgpt}, as well as ChatGLM~\cite{glm_130b}, which is specifically trained on Chinese corpora.
ChatGLM~\cite{glm_130b} achieves token-level accurate results but lacks diversity in its generated texts. Thus, ChatGPT-4~\cite{openai2023chatgpt} is adopted.

However, for the PHOENIX14T, the German sign language dataset, we have not found open-source powerful LLMs tailored to German can produce satisfactory references. Therefore, although ChatGPT-4 is relatively costly, it could potentially save the cost of sentence rephrasing by native speakers due to its high-performance in German sentence generation. 

\section{Details of Human Verification and Modification} 
In our human verification and modification process, we invite 20 native speakers of German and Chinese, respectively, to review and modify the raw/original language multi-references generated by the LLM. The revision guidelines focus on both correcting grammatical errors and preserving the meaning. Each instance is annotated by multiple people through a cross-check verification process. Specifically, we ask each native speaker as an examiner to cross-check around 25\% of randomly-chosen data labeled by another annotator. If more than 10\% of the annotated data have obvious errors, a third annotator will review and correct the annotations.

\begin{table*}[t]
\tiny
\centering
\caption{An example (from CSL-Daily~\cite{CSL_Daily}) with multiple references generated by different LLMs.} 
\vspace{-1em}
\label{multi_ref_case}
\begin{tabularx}{0.95\textwidth}{l|c|c}
\toprule
\multirow{1}{*}{\tiny{Source}} & \multicolumn{2}{l}{\adjustbox{valign=m}{\includegraphics[width=0.75\textwidth]{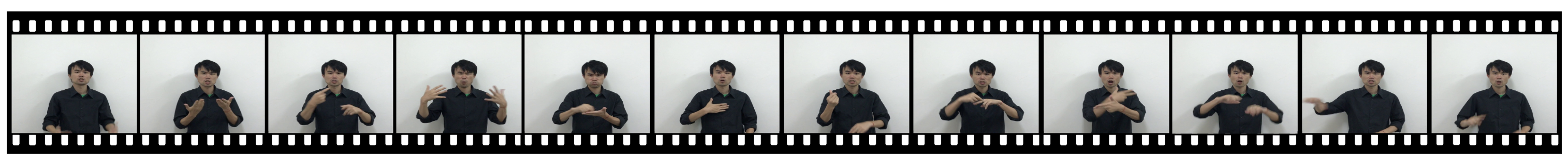}}} \\
\midrule
{\tiny{Target}} & \multicolumn{2}{l}{
\begin{CJK}{UTF8}{gbsn}\tiny{为了争取全额奖学金，我写了N遍申请书。}\end{CJK} \tiny{(\textbf{EN:} In order to get a full scholarship, I wrote applications N times.) }} \\ \midrule
\multicolumn{1}{l|}{\tiny{LLM Model}} & \multicolumn{1}{c|}{\tiny{ChatGLM~\cite{glm_130b}}} &  \tiny{ChatGPT-4~\cite{openai2023chatgpt}} \\ \midrule
{\tiny{Prompt 1}} & \multicolumn{2}{l}{
\begin{CJK}{UTF8}{gbsn}\tiny{请帮我提取\textbf{~Target~}中的关键词。}\end{CJK} \tiny{(\textbf{EN:} Please help me extract the keywords from the \textbf{Target}.) }} \\ \midrule
\multicolumn{1}{l|}{\multirow{2}{*}{\tiny{Keywords}}} & \multicolumn{1}{c|}{\begin{CJK}{UTF8}{gbsn}\tiny{争取，全额，奖学金，我，写，N遍，申请书}\end{CJK}} & \begin{CJK}{UTF8}{gbsn}\tiny{为了，争取，全额奖学金，写，N遍，申请书}\end{CJK} \\ 
\multicolumn{1}{l|}{} & \multicolumn{1}{c|}{strive, all, scholarship, I, write, N times, application} & for, strive, full scholarship, write, N times, application \\ \midrule
\multirow{3}{*}{\tiny{Prompt 2}} & \multicolumn{2}{l}{
\begin{CJK}{UTF8}{gbsn}\tiny{请你结合提取到的关键词~\textbf{Keywords}~和原句~\textbf{Target}~,帮我生成一些与原句相似的句子。注意关键词不是必选。}\end{CJK}} \\
 & \multicolumn{2}{l}{\tiny{(\textbf{EN:} Please combine the extracted keywords, \textbf{Keywords}, with the original sentence, \textbf{Target}, help me generate some sentences similar to the original.}} \\ 
& \multicolumn{2}{l}{\tiny{Note that the keywords are not mandatory.)}} \\\midrule
\multicolumn{1}{l|}{\multirow{7}{*}{\tiny{Multi-Reference}}} & \multicolumn{1}{l|}{
\begin{CJK}{UTF8}{gbsn}为获得全额奖学金，我写了多次申请书。\end{CJK}
} & 
\multicolumn{1}{l}{\begin{CJK}{UTF8}{gbsn}为了我能得到全额奖学金，我反复撰写了多次申请书。\end{CJK}} \\
\multicolumn{1}{l|}{} & \multicolumn{1}{l|}{I wrote multiple applications to get a full scholarship.}  & \multicolumn{1}{l}{In order to get a full scholarship, I wrote multiple applications.}\\
\multicolumn{1}{l|}{} & \multicolumn{1}{l|}{
\begin{CJK}{UTF8}{gbsn}为了争取全部奖学金，我写了多封申请书。\end{CJK}
} &  \multicolumn{1}{l}{\begin{CJK}{UTF8}{gbsn}我为能竞取全额奖学金，写下了N份申请书。\end{CJK}} \\
\multicolumn{1}{l|}{} & \multicolumn{1}{l|}{In order to win the full scholarship, I wrote multiple applications.~~~~~~} & \multicolumn{1}{l}{In order to compete for a full scholarship, I wrote N applications.
} \\
\multicolumn{1}{l|}{} & \multicolumn{1}{l|}{
\begin{CJK}{UTF8}{gbsn}我为争取全额奖学而写了多次申请书。\end{CJK}
} & \multicolumn{1}{l}{\begin{CJK}{UTF8}{gbsn}我争取全额奖学金，写了一次又一次的申请书。\end{CJK}} \\
\multicolumn{1}{l|}{} & \multicolumn{1}{l|}{I wrote multiple applications to get a full scholarship.}  & \multicolumn{1}{l}{I fought for a full scholarship and wrote applications again and again.} \\
\midrule
\multicolumn{1}{l|}{\tiny{Evaluation (k=3)}} & \multicolumn{1}{c|}{\tiny{Ave.rfb~\textbackslash~Ave.pwb~\textbackslash~Ave.rfbRT~\textbackslash~Human=~\textbf{54.19}~\textbackslash~44.48~\textbackslash~80.77~\textbackslash~2}} & \tiny{Ave.rfb~\textbackslash~Ave.pwb~\textbackslash~Ave.rfbRT~\textbackslash~Human=~44.70~\textbackslash~~\textbf{37.37}~\textbackslash~~\textbf{82.01}~\textbackslash~~\textbf{1}} \\
\bottomrule
\end{tabularx}
\vspace{-1em}
\end{table*}

\section{Main Results on Validation Set}
Following~\citet{NSLT}, in Table~\ref{DSLT_DE_Dev} and Table~\ref{DSLT_ZH_Dev}, we also report the performance on the \textit{validation} set.
The results indicate that our DivSLT model, trained with multiple references, exhibits improvements on both translation accuracy and diversity.
Specifically, benefiting from the reinforcement learning method, the DivSLT model improves the B-BM score and the BLEURT-BM score by approximately +3 (21.17$\rightarrow$24.29) and +11 (53.91$\rightarrow$64.70) on the PHOENIX14T.
As for the CSL-Daily dataset, the B-BM score and the BLEURT-BM score are enhanced by about +4.5 (8.49$\rightarrow$13.11) and +9 (34.51$\rightarrow$43.52).
Most importantly, the results of the Top-3 generated translations from the DivSLT surpass those from the GFSLT across the majority of diversity and accuracy evaluation metrics.


\section{DivSLT with Vision-Language Pre-training Strategies}
\begin{figure}[h]
  \centering
  \includegraphics[width=1\linewidth]{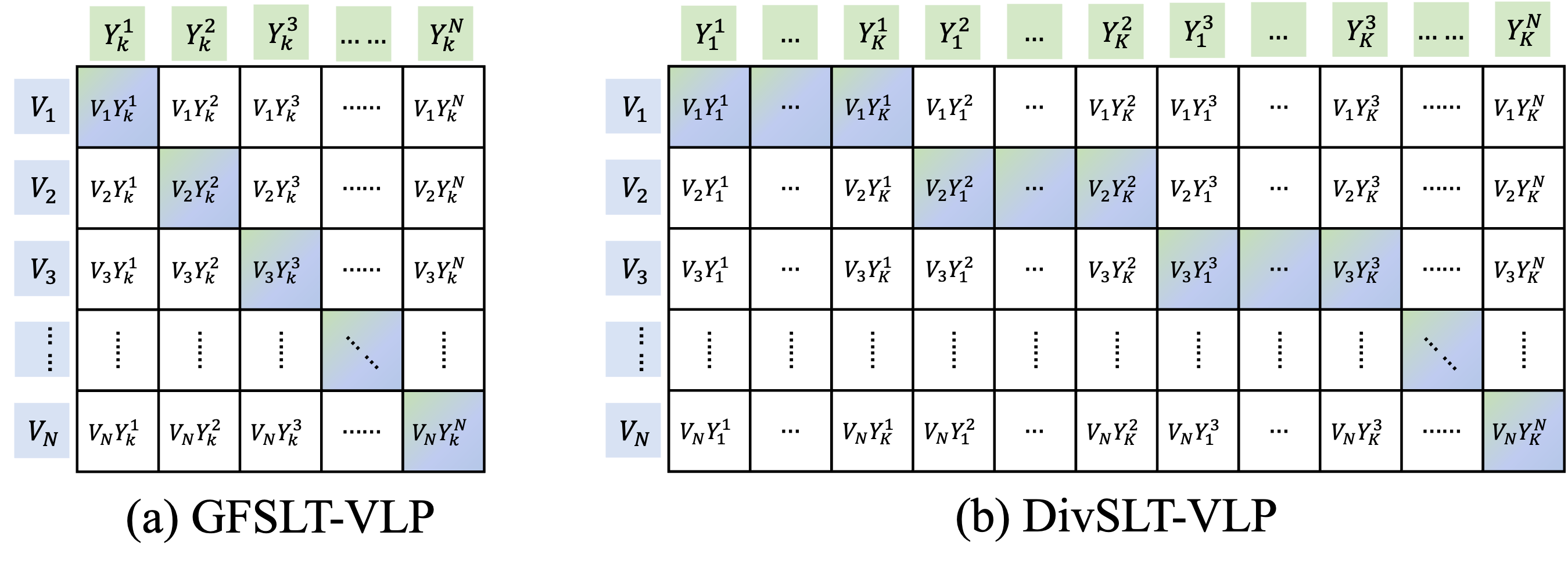}
  \caption{Different visual-language pre-training strategies. (a) One-to-One GFSLT-VLP~\cite{GF_VLP}. (b) One-to-Many DivSLT-VLP.}
  \label{div-vlp}
\end{figure}

\noindent To improve the sign language translation performance, \citet{GF_VLP} propose a vision-language pre-training (VLP) method that integrates CLIP~\cite{clip} with the masked self-supervised learning to narrow down the semantic gap between visual and textual representations and reconstruct masked sentences.
{
\setlength{\tabcolsep}{5.5pt}
\begin{table*}[!t]
\centering
\tiny
\caption{Experimental results on the \textbf{validation} set of PHOENIX14T-Div. ``BM'', ``MR'', ``R'', ``B'' and ``BRT'' represent best-matching, multi-reference, ROUGE, BLEU and BLEURT. $^*$ represents our reproduced results.}
\vspace{-1em}
\label{DSLT_DE_Dev}
\begin{tabular}{@{}l|ccccc|cccccc@{}} \toprule
 & \multicolumn{5}{c|}{Top-3 Predictions} & \multicolumn{6}{c}{Top-1 Prediction} \\ \midrule
{\bf Single-Reference SLT} & rfb-BM~$\uparrow$ & mrfb~$\uparrow$ & \textbf{\textcolor{blue}{pwb}}~$\downarrow$ & rfbRT-BM~$\uparrow$ & mrfbRT~$\uparrow$ & BRT-MR~$\uparrow$ & R-MR~$\uparrow$ & B-MR~$\uparrow$ & \textbf{\textcolor{blue}{BRT-BM}}~$\uparrow$ & R-BM~$\uparrow$ & B-BM~$\uparrow$ \\ \midrule 
GFSLT~\cite{GF_VLP}$^*$ & 20.22 & 21.66 & 38.98 & 51.56 & 43.96 & 46.09 & 44.29 & 22.87  & 53.91&	42.79&	21.17 \\ \midrule \midrule
{\bf Multi-Reference SLT} & rfb-BM~$\uparrow$ & mrfb~$\uparrow$ & \textbf{\textcolor{blue}{pwb}}~$\downarrow$ & rfbRT-BM~$\uparrow$ & mrfbRT~$\uparrow$ & BRT-MR~$\uparrow$ & R-MR~$\uparrow$ & B-MR~$\uparrow$ & \textbf{\textcolor{blue}{BRT-BM}}~$\uparrow$ & R-BM~$\uparrow$ & B-BM~$\uparrow$ \\ \midrule 
DivSLT [Stage 1] & 19.88 & 23.33 & \textbf{22.73} & \textbf{61.19} & 54.01  & \textbf{57.78}& 48.75 & 27.35 & 64.09&	46.62&	23.70 \\
DivSLT [Stage 2] & \textbf{21.62} & \textbf{25.08} & 26.65 & 60.88 & \textbf{54.82} &  57.08   & \textbf{49.70} & \textbf{28.15} & \textbf{64.70}  & \textbf{47.60} & \textbf{24.29} \\ \bottomrule
\end{tabular}
\vspace{-1em}
\end{table*}
}
{
\setlength{\tabcolsep}{5.5pt}
\begin{table*}[!t]
\centering
\tiny
\caption{Experimental results on the \textbf{validation} set of CSL-Daily-Div. ``BM'', ``MR'', ``R'', ``B'' and ``BRT'' represent best-matching, multi-reference, ROUGE, BLEU and BLEURT, respectively. $^*$ represents our reproduced results.}
\vspace{-1em}
\label{DSLT_ZH_Dev}
\begin{tabular}{@{}l|ccccc|cccccc@{}} \toprule
 & \multicolumn{5}{c|}{Top-3 Predictions} & \multicolumn{6}{c}{Top-1 Prediction} \\ \midrule
{\bf Single-Reference SLT} & rfb-BM~$\uparrow$ & mrfb~$\uparrow$ & \textbf{\textcolor{blue}{pwb}}~$\downarrow$ & rfbRT-BM~$\uparrow$ & mrfbRT~$\uparrow$ & BRT-MR~$\uparrow$ & R-MR~$\uparrow$ & B-MR~$\uparrow$ & \textbf{\textcolor{blue}{BRT-BM}}~$\uparrow$ & R-BM~$\uparrow$ & B-BM~$\uparrow$ \\ \midrule 
GFSLT~\cite{GF_VLP}$^*$  & 9.23 & 11.91 & 39.72 & 28.56 & 27.84 & 32.98 & 33.96 & 11.31 & 34.51  & 33.39 & 8.49  \\ \midrule \midrule
{\bf Multi-Reference SLT} & rfb-BM~$\uparrow$ & mrfb~$\uparrow$ & \textbf{\textcolor{blue}{pwb}}~$\downarrow$ & rfbRT-BM~$\uparrow$ & mrfbRT~$\uparrow$ & BRT-MR~$\uparrow$ & R-MR~$\uparrow$ & B-MR~$\uparrow$ & \textbf{\textcolor{blue}{BRT-BM}}~$\uparrow$ & R-BM~$\uparrow$ & B-BM~$\uparrow$ \\ \midrule 
DivSLT [Stage 1] & 11.60 & 13.73 & \textbf{23.36}  & 40.04 & 33.54 & 35.98 & 34.73 & 14.27 & 42.72& 	32.18& 	12.12 \\
DivSLT [Stage 2] & \textbf{12.85} & \textbf{15.03} & 26.59 &\textbf{41.52} & \textbf{35.12} & \textbf{36.82} &  \textbf{36.25} & \textbf{15.37} & \textbf{43.52} & \textbf{33.97} & \textbf{13.11}  \\ \bottomrule
\end{tabular}
\vspace{-1em}

\end{table*}
}
{
\setlength{\tabcolsep}{4.9pt}
\begin{table*}[!ht]
\centering
\tiny
\caption{Experimental results of DivSLT with vision-language pre-training on the \textbf{test} set of PHOENIX14T-Div. ``BM'', ``MR'', ``R'', ``B'' and ``BRT'' represent best-matching, multi-reference, ROUGE, BLEU and BLEURT, respectively. $^*$ represents our reproduced results.}
\vspace{-1em}
\label{DSLT_De_with_VLP}
\begin{tabular}{@{}l|ccccc|cccccc@{}} \toprule
 & \multicolumn{5}{c|}{Top-3 Predictions} & \multicolumn{6}{c}{Top-1 Prediction} \\ \midrule
{\bf Single-Reference SLT} & rfb-BM~$\uparrow$ & mrfb~$\uparrow$ & \textbf{\textcolor{blue}{pwb}}~$\downarrow$ & rfbRT-BM~$\uparrow$ & mrfbRT~$\uparrow$ & BRT-MR~$\uparrow$ & R-MR~$\uparrow$ & B-MR~$\uparrow$ & \textbf{\textcolor{blue}{BRT-BM}}~$\uparrow$ & R-BM~$\uparrow$ & B-BM~$\uparrow$ \\ \midrule 
GFSLT~\cite{GF_VLP}$^*$  & 19.03 & 20.41 & 40.86 & 51.73 & 43.82  & 46.45 & 42.53 & 21.29 & 54.70 & 41.14 & 19.76 \\ 
GFSLT-VLP~\cite{GF_VLP}$^*$ & 21.11 & 22.74 & 40.93 & 54.23 & 45.93 & 48.09  & 45.36 & 23.89 & 56.56&	43.90&	22.02 \\ \midrule \midrule
{\bf Multi-Reference SLT} & rfb-BM~$\uparrow$ & mrfb~$\uparrow$ & \textbf{\textcolor{blue}{pwb}}~$\downarrow$ & rfbRT-BM~$\uparrow$ & mrfbRT~$\uparrow$ & BRT-MR~$\uparrow$ & R-MR~$\uparrow$ & B-MR~$\uparrow$ & \textbf{\textcolor{blue}{BRT-BM}}~$\uparrow$ & R-BM~$\uparrow$ & B-BM~$\uparrow$ \\ \midrule 
GFSLT-VLP [Stage 1] & 19.67 & 22.81 & 22.45 & 60.67 & 53.03  & 56.80 & 47.54 & 26.87 & 64.21&	45.30&	23.21 \\
GFSLT-VLP [Stage 2]& 21.08 & 24.36 & 26.26 & 60.34 & 53.98  & 56.09 & 48.95 & 27.25  & 64.17&	46.54&	23.80\\ \midrule
DivSLT [Stage 1] & 20.74 & 23.96 & 22.26 & \textbf{61.89} & 54.14 & 57.15 & 48.24 & 27.69 & 64.82&	46.02&	24.20  \\
DivSLT [Stage 2] & \textbf{22.12} & 25.23 & 26.30 & 61.34 & 54.97& 58.02 & 49.82 & 28.82 & 64.95&	47.61&	25.29
\\ \midrule
DivSLT-VLP [Stage 1] & 20.12 & 24.31 & \textbf{22.07} & 61.14 & 54.41  & 57.56 & 48.77 & 27.94 & 65.04&	46.27&	24.13\\
DivSLT-VLP [Stage 2]  &21.78 & \textbf{26.57} & 25.56 & 61.56 & \textbf{55.19}  & \textbf{58.81} & \textbf{50.51} & \textbf{29.26}& \textbf{65.69} & \textbf{48.01}	&\textbf{25.79}  \\ \bottomrule
\end{tabular}
\vspace{-2em}
\end{table*}
}
{
\setlength{\tabcolsep}{4.8pt}
\begin{table*}[t]
\centering
\tiny
\caption{\XS{Benchmarking results for gloss-based DivSLT on PHOENIX14T-Div. ``BM'', ``MR'', ``R'', ``B'' and ``BRT'' represent best-matching, multi-reference, ROUGE, BLEU and BLEURT, respectively.}}
\vspace{-1em}
\label{gloss_based_phx14t}
\begin{tabular}{@{}l|ccccc|cccccc@{}} \toprule
 & \multicolumn{5}{c|}{Top-3 Predictions} & \multicolumn{6}{c}{Top-1 Prediction} \\ \midrule
{\bf Single-Reference Gloss-Based SLT} & rfb-BM~$\uparrow$ & mrfb~$\uparrow$ & \textbf{\textcolor{blue}{pwb}}~$\downarrow$ & rfbRT-BM~$\uparrow$ & mrfbRT~$\uparrow$ & BRT-MR~$\uparrow$ & R-MR~$\uparrow$ & B-MR~$\uparrow$ & \textbf{\textcolor{blue}{BRT-BM}}~$\uparrow$ & R-BM~$\uparrow$ & B-BM~$\uparrow$ \\ \midrule 
MMTLB~\cite{mmtlb} &24.83&	28.84&	61.39&	60.21&	50.05& 52.60&	51.25&	29.16&	59.94&	52.65&	28.39     \\
TS-SLT~\cite{two_s} & \textbf{26.34} &	\textbf{29.61}& 60.92& 	\textbf{61.76}&	52.54& 53.19&	\textbf{52.56}&	\textbf{31.06}&	62.40&	\textbf{53.48}&	\textbf{28.95}    \\ \midrule
{\bf Multi-Reference Gloss-Free SLT} & rfb-BM~$\uparrow$ & mrfb~$\uparrow$ & \textbf{\textcolor{blue}{pwb}}~$\downarrow$ & rfbRT-BM~$\uparrow$ & mrfbRT~$\uparrow$ & BRT-MR~$\uparrow$ & R-MR~$\uparrow$ & B-MR~$\uparrow$ & \textbf{\textcolor{blue}{BRT-BM}}~$\uparrow$ & R-BM~$\uparrow$ & B-BM~$\uparrow$ \\ \midrule 
DivSLT & 22.12 & 25.23 & \textbf{26.30} & 61.34 & \textbf{54.97}  &  \textbf{58.02} & 49.82 & 28.82 & \textbf{64.95}&	47.61&	25.29 \\  \bottomrule
\end{tabular}
\vspace{-1em}
\end{table*}
}
{
\setlength{\tabcolsep}{4.8pt}
\begin{table*}[t]
\centering
\tiny
\caption{\XS{Benchmarking results for gloss-based DivSLT on CSL-Daily-Div.``BM'', ``MR'', ``R'', ``B'' and ``BRT'' represent best-matching, multi-reference, ROUGE, BLEU and BLEURT, respectively.}}
\vspace{-1em}
\label{gloss_based_csl}
\begin{tabular}{@{}l|ccccc|cccccc@{}} \toprule
{\bf Single-Reference Gloss-Based SLT} & rfb-BM~$\uparrow$ & mrfb~$\uparrow$ & \textbf{\textcolor{blue}{pwb}}~$\downarrow$ & rfbRT-BM~$\uparrow$ & mrfbRT~$\uparrow$ & BRT-MR~$\uparrow$ & R-MR~$\uparrow$ & B-MR~$\uparrow$ & \textbf{\textcolor{blue}{BRT-BM}}~$\uparrow$ & R-BM~$\uparrow$ & B-BM~$\uparrow$ \\ \midrule 
MMTLB ~\cite{mmtlb} &22.51&	26.52&	59.75&	50.77&	35.14& 35.60&	33.26&	28.51&	53.03&	35.25&	21.92     \\
TS-SLT ~\cite{two_s} &\textbf{23.78}&	\textbf{28.85}&	60.49&	\textbf{54.86}&	\textbf{36.69}& 36.40&	33.75&	\textbf{29.20}&	\textbf{54.65}&	\textbf{37.72}&	\textbf{23.79}    \\ \midrule
{\bf Multi-Reference Gloss-Free SLT} & rfb-BM~$\uparrow$ & mrfb~$\uparrow$ & \textbf{\textcolor{blue}{pwb}}~$\downarrow$ & rfbRT-BM~$\uparrow$ & mrfbRT~$\uparrow$ & BRT-MR~$\uparrow$ & R-MR~$\uparrow$ & B-MR~$\uparrow$ & \textbf{\textcolor{blue}{BRT-BM}}~$\uparrow$ & R-BM~$\uparrow$ & B-BM~$\uparrow$ \\ \midrule
DivSLT & 13.24 & 15.38 &\textbf{ 26.50} &41.70 & 35.31 & \textbf{36.42} & \textbf{35.23} & 15.24 & 43.01&32.74&13.03  \\  \bottomrule
\end{tabular}
\vspace{-2em}
\end{table*}
}
{
\setlength{\tabcolsep}{4.8pt}
\begin{table*}[t]
\centering
\tiny
\caption{\XS{The performance of GFSLT + ChatGPT-4 rewriting on PHOENIX14T-Div. ``BM'', ``MR'', ``R'', ``B'' and ``BRT'' represent best-matching, multi-reference, ROUGE, BLEU and BLEURT, respectively. $^*$ represents our reproduced results.}}
\vspace{-1em}
\label{llm_rewriting}
\begin{tabular}{@{}l|ccccc|cccccc@{}} \toprule
& \multicolumn{5}{c|}{Top-3 Predictions} & \multicolumn{6}{c}{Top-1 Prediction} \\ \midrule
& rfb-BM~$\uparrow$ & mrfb~$\uparrow$ & \textbf{\textcolor{blue}{pwb}}~$\downarrow$ & rfbRT-BM~$\uparrow$ & mrfbRT~$\uparrow$ & BRT-MR~$\uparrow$ & R-MR~$\uparrow$ & B-MR~$\uparrow$ & \textbf{\textcolor{blue}{BRT-BM}}~$\uparrow$ & R-BM~$\uparrow$ & B-BM~$\uparrow$ \\ \midrule 
GFSLT$^*$  & 19.03 & 20.41 & 40.86 & 51.73 & 43.82  & 46.45 & 42.53 & 21.29 & 54.70 & 41.14 & 19.76     \\
GFSLT$^*$ + ChatGPT-4  & 11.38&	13.32&	\textbf{19.75}&	55.96&	50.15& 50.82&	34.96&	14.64&	56.85&	32.55&	12.51   \\ 
DivSLT & \textbf{22.12} & \textbf{25.23} & 26.30 & \textbf{61.34} & \textbf{54.97}  &  \textbf{58.02} & \textbf{49.82} & \textbf{28.82} & \textbf{64.95}&	\textbf{47.61}&	\textbf{25.29} \\
\bottomrule
\end{tabular}
\vspace{-1em}
\end{table*}
}

{
\setlength{\tabcolsep}{4.8pt}
\begin{table*}[t]
\centering
\tiny
\caption{\XS{Benchmarking results for gloss-based DivSLT on How2Sign-Div.``BM'', ``MR'', ``R'', ``B'' and ``BRT'' represent best-matching, multi-reference, ROUGE, BLEU and BLEURT, respectively.}}
\vspace{-1em}
\label{how2sign}
\begin{tabular}{@{}l|ccccc|cccccc@{}} \toprule
{\bf Single-Reference SLT} & rfb-BM~$\uparrow$ & mrfb~$\uparrow$ & \textbf{\textcolor{blue}{pwb}}~$\downarrow$ & rfbRT-BM~$\uparrow$ & mrfbRT~$\uparrow$ & BRT-MR~$\uparrow$ & R-MR~$\uparrow$ & B-MR~$\uparrow$ & \textbf{\textcolor{blue}{BRT-BM}}~$\uparrow$ & R-BM~$\uparrow$ & B-BM~$\uparrow$ \\ \midrule 
GFSLT ~\cite{GF_VLP} & 10.92&	11.94&	38.05&		29.16&	24.25&	29.60&	32.35&	10.52&		36.75&	27.87&	9.14\\ \midrule
{\bf Multi-Reference SLT} & rfb-BM~$\uparrow$ & mrfb~$\uparrow$ & \textbf{\textcolor{blue}{pwb}}~$\downarrow$ & rfbRT-BM~$\uparrow$ & mrfbRT~$\uparrow$ & BRT-MR~$\uparrow$ & R-MR~$\uparrow$ & B-MR~$\uparrow$ & \textbf{\textcolor{blue}{BRT-BM}}~$\uparrow$ & R-BM~$\uparrow$ & B-BM~$\uparrow$ \\ \midrule
DivSLT & \textbf{13.12}&	\textbf{16.47}&	\textbf{27.58}&		\textbf{38.13}&	3\textbf{4.21}&	\textbf{35.57}&	\textbf{34.77}&	\textbf{13.51}&		\textbf{41.96}&	\textbf{29.06}&	\textbf{10.03}   \\  \bottomrule
\end{tabular}
\vspace{-2em}
\end{table*}
}

However, the larger vocabulary size of PHOENIX14T-Div (9,427 words) prohibits the direct use of VLP models trained on PHOENIX14T with smaller vocabulary size (2,887 words).
To address this issue, we conduct pre-training on PHOENIX14T-Div to adapt VLP to a larger vocabulary size.
As demonstrated in Table~\ref{DSLT_De_with_VLP}, VLP can effectively improve the Top-1 prediction accuracy of GFSLT (B-BM \& R-BM).

However, VLP~\cite{GF_VLP} does not improve the performance of DivSLT, and DivSLT even outperforms GFSLT-VLP in terms of accuracy (B-BM: 22.02~\textit{vs.}~25.29), diversity (pwb: 40.93~\textit{vs.}~26.30) and semantic precision (BRT-BM: 56.56~\textit{vs.}~64.95).
We speculate that VLP is tailored for the one-to-one SLT rather than one-to-many mappings, and thus a pre-training strategy designed for diverse SLT is also desirable.

Inspired by the one-to-one VLP strategy, as depicted in Figure~\ref{div-vlp}(b), we introduce a simple yet effective pre-training strategy for DivSLT.
We conduct pre-training for 80 epochs on PHOENIX14T-Div.
The results in Table~\ref{DSLT_De_with_VLP} demonstrate that our proposed pre-training tailored for DivSLT enhances translation accuracy, diversity and semantic precision. 
This implies that a better pre-training for DivSLT can mitigate issues caused by a larger vocabulary size of multi-references.




\section{Quality of the top-\textit{k}th translation.}
\begin{figure}[t]
\begin{center}
\includegraphics[width=1\linewidth]{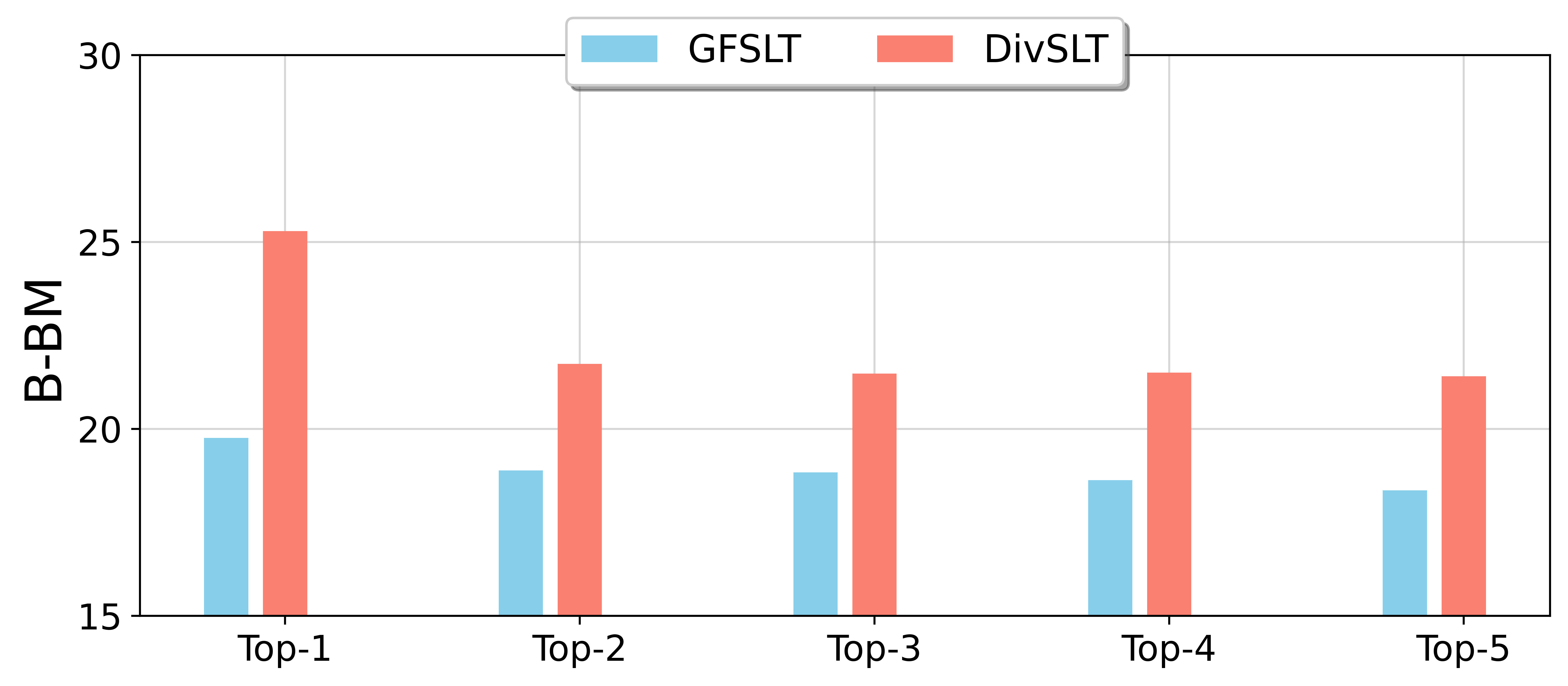}
\end{center}
\vspace{-1.5em}
\caption{B-BM scores of the Top-\(k\)th hypothesis generated by GFSLT and DivSLT.} 
\label{topk-b}
\end{figure}

\begin{figure}[t]
\begin{center}
\includegraphics[width=1\linewidth]{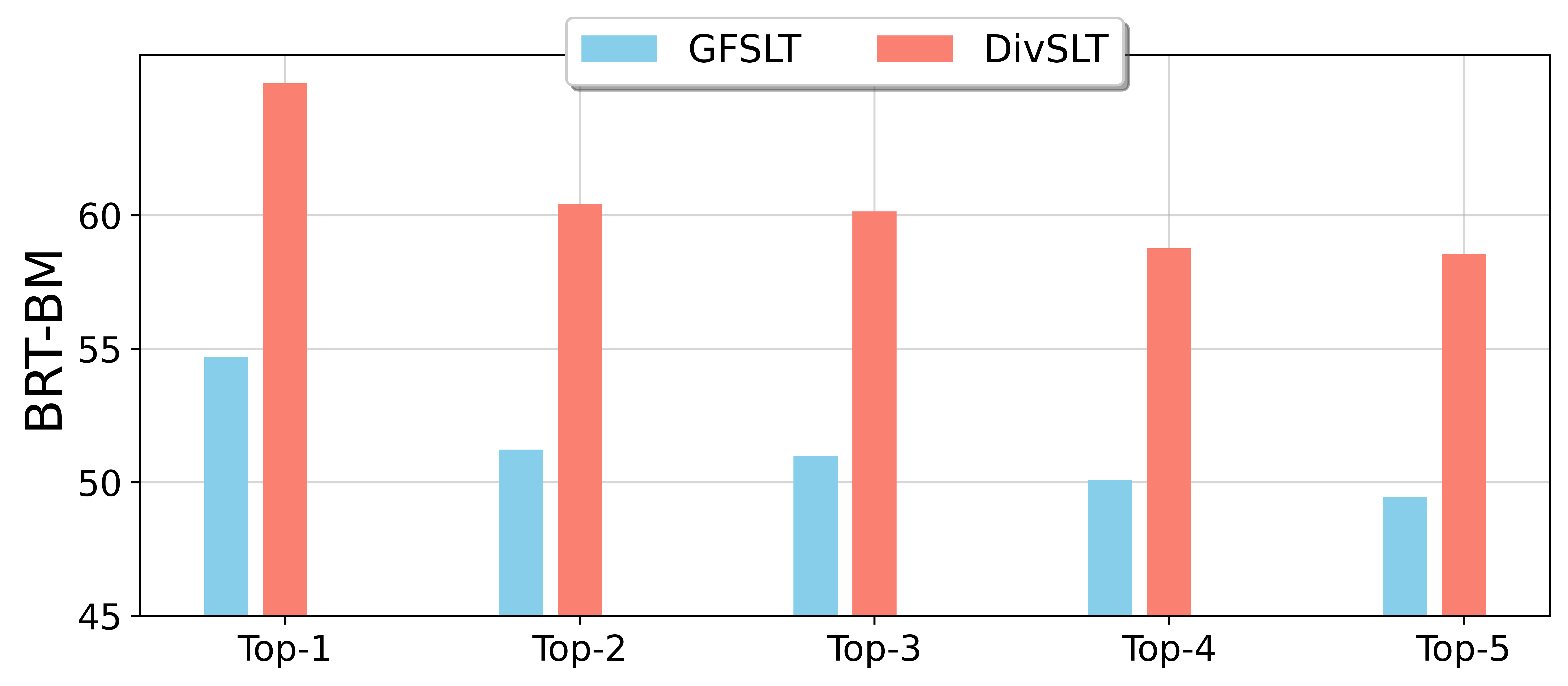}
\end{center}
\vspace{-1.5em}
\caption{BRT-BM scores of the Top-\(k\)th hypothesis generated by GFSLT and DivSLT.} 
\label{topk-brt}
\end{figure}

Figure~\ref{topk-b} and Figure~\ref{topk-brt} depict the accuracy (B-BM) and the semantic similarity metric (BRT-BM) for translation results generated by diverse beam search on the PHOENIX14T-Div. 
It is evident that the accuracy and semantic precision of the Top-kth translations from DivSLT consistently surpass those from GFSLT.
This indicates that DivSLT not only translates with higher overall accuracy but also maintains closer semantic alignment with the target language.


\section{\XS{Benchmarking for gloss-based DivSLT}}
\XS{We include some gloss-based SLT methods in our DivSLT benchmark. To the best of our knowledge, MMTLB~\cite{mmtlb} and TwoStream-SLT (TS-SLT)~\cite{two_s} are the state-of-the-art open-sourced gloss-based SLT approaches, and TS-SLT also achieves the best performance on the CSL-Daily and PHOENIX14T datasets among the open-sourced gloss-based SLT models.}

\XS{Table~\ref{gloss_based_phx14t} and Table~\ref{gloss_based_csl} demonstrate that with gloss supervision, gloss-based models significantly outperform gloss-free models on n-gram based metrics. These improvements are primarily attributed to the gloss annotations, which significantly facilitate the correspondence establishment between sign gestures and words. However, our DivSLT model outperforms the gloss-based model in terms of diversity, indicated by pwb, and is close to or even superior to the gloss-based models in semantic similarity assessments, denoted by BRT-MR. It indicates that the translations generated by DivSLT are diverse and semantically accurate. }

\section{\XS{DivSLT on How2Sign}}
To explore the performance of the DivSLT model on larger datasets, we select How2Sign~\cite{how2sign}. 
Compared to other larger SLT datasets, How2Sign is not only rich in data content but also sufficiently clean. 
We generate multi-references using the method mentioned in Section 3. 
It is important to note that no manual verification for How2Sign. 
As shown in Table~\ref{how2sign}, the results using DivSLT surpass those of GFSLT in terms of accuracy, diversity, and semantic precision. 
Additionally, even without manual verification, it is still possible to enhance the diversity of SLT model translations using LLMs.

\section{\XS{DivSLT with LLM rewriting}}
\XS{We explore improving the accuracy and diversity of single-reference SLT models through LLM rewriting.
We take the PHOENIX14T-Div dataset as an example. To be specific, we first employ ChatGPT-4 to rephrase the translated results of GFSLT~\cite{GF_VLP} by first examining whether a sentence is fluent or not. If a sentence is unfluent, ChatGPT-4 will rephrase the sentence with minimal revision. Then, we further leverage ChatGPT-4 to generate diverse translations based on the previous refined translation results. 
Table~\ref{llm_rewriting} indicates that using ChatGPT-4 to rewrite translation results of GFSLT can improve the diversity of translations. However, GFSLT + ChatGPT-4 suffers a significant drop in n-gram evaluation metrics. In comparison, our DivSLT model still outperforms the variant GFSLT + ChatGPT-4 on most of the evaluation metrics.}

\section{Case Study}
More case studies from the extended datasets are shown in Table \ref{more_case_study}.
The results generated by our DivSLT are highly similar to the ground-truth reference in semantics but different in word usage.
In the penultimate example from CSL-Daily-Div, both ``\begin{CJK}{UTF8}{gbsn}爱护\end{CJK}'' (care) and ``\begin{CJK}{UTF8}{gbsn}关爱\end{CJK}'' are acceptable in Chinese to express the concept of ``taking care of somebody or something''.
In the last example from CSL-Daily-Div, we find that the translation results of DivSLT are roughly similar in meaning to the ground truth. 
However, there are still some untranslated elements, \textit{e.g.}, ``\begin{CJK}{UTF8}{gbsn}肯定\end{CJK}'', and instances of over-translation, \textit{e.g.}, ``\begin{CJK}{UTF8}{gbsn}们\end{CJK}''.
Overall, diverse sign language translation facilitates better communication between hearing-impaired individuals and the hearing community.

\section{\XS{Practicality and Application Prospects of DivSLT}}
\XS{Since a single sign video could correspond to multiple plausible translations rather than the single ground-truth reference provided by a sign language dataset and existing SLT metrics do not support the evaluation of diverse translated results, the understandability of the translation results cannot be fully reflected. In contrast, the application of DivSLT is able to overcome these limitations and facilitates the interpretation of SLT results. As indicated by the results in Table 3 and Table 4 (the main paper), we can see that translation results (evaluated by multi-reference based metrics) are actually better (i.e., easier to understand) than expected when they are evaluated by single references. This demonstrates that DivSLT can actually improve the understandability of translated sentences.}

\XS{Moreover, when the number of sign videos is limited due to the difficulty of attainment of sign annotations, enriching spoken language supervision data is an alternative solution to achieving better translation performance. With more spoken language supervision, DivSLT can achieve better generalization abilities in producing diverse yet more fluent translations, as indicated by the BRT-BM scores in Table 3 and Table 4. Specifically, when our DivSLT model is trained with multi-references, the B-BM and BRT-BM scores are higher. This implies that the translation results achieve better readability. Therefore, we believe that DivSLT can be used in any application related to SLT.}


\clearpage
\begin{table*}[t]
\centering
\tiny
\caption{Case Study. We highlight the differences between sentences. \textcolor{myred}{Red} indicates agreement with the references. \textcolor{myblue}{Blue} denotes correct but different words.  \textcolor{mygray}{Gray} represents missing contents. }
\vspace{-1em}
\label{more_case_study}
\begin{tabular}{p{\linewidth}} \toprule 
\textbf{Ground Truth (PHOENIX14T):} heute nacht ist es im süden teils klar teilweise aber auch nebel \\
(EN: Tonight in the south, it is partly clear, but there is also fog at times.) \\  \midrule
\textbf{Extended Multiple References (PHOENIX14T-Div):} \\ 
1. heute abend wird es im süden zumeist klar sein jedoch teilweise auch nebelig . \\ 
(EN: This evening in the south, it will be mostly clear, but there will also be patches of fog.) \\
2. es wird heute in der nacht im süden teilweise klar aber auch etwas nebelig sein\\
(EN: Tonight in the south, there will be periods of clear weather, but also some fog.) \\
3. heute nacht gibt es im süden eine mischung aus klarem himmel und teils nebel \\
(EN: Tonight in the south, there will be a mix of clear skies and some fog.) \\
\midrule
\textbf{Top-3 Predictions:}\\
1. \textcolor{myred}{heute nacht ist es im süden} \textcolor{myblue}{zum teil} \textcolor{myred}{klar} hin \textcolor{myred}{und} \textcolor{mygray}{(aber)} \textcolor{myblue}{wieder bildet sich} \textcolor{myred}{nebel}\\
(EN: \textcolor{myred}{Tonight in the south}, \textcolor{myred}{it is} \textcolor{myblue}{partly} \textcolor{myred}{clear}, and \textcolor{mygray}{(but)} \textcolor{myred}{fog} \textcolor{myblue}{will form now and then}.) \\
2. \textcolor{myred}{heute nacht ist der himmel im süden klar} \textcolor{myblue}{mit einigen nebelfeldern zu sehen} \textcolor{myred}{aber auch} \textcolor{myblue}{vereinzelt bildet sich} \textcolor{myred}{nebel}\\
(EN: \textcolor{myred}{Tonight the sky in the south is clear} \textcolor{myblue}{with some patches of fog}, \textcolor{myred}{but some fog} \textcolor{myblue}{is also forming}.) \\
3. \textcolor{myred}{im süden ist es heute nacht} \textcolor{myblue}{zum teil} \textcolor{myred}{klar} \textcolor{mygray}{(aber)} \textcolor{myblue}{ab und zu bildet sich} \textcolor{myred}{nebel}\\
(EN: \textcolor{myred}{Tonight in the south, it will be} \textcolor{myblue}{partly} \textcolor{myred}{clear} with \textcolor{mygray}{(but)} \textcolor{myred}{fog} \textcolor{myblue}{forming from time to time}.) \\ \midrule \midrule

\textbf{Gound Truth (PHOENIX14T):} morgen scheint dann wieder überall die sonne .\\
(EN: Tomorrow the sun will shine everywhere again.) \\  \midrule
\textbf{Extended Multiple References (PHOENIX14T-Div):} \\ 
1. morgen wird die sonne wieder überall scheinen . \\ 
(EN: Tomorrow the sun will shine all over again.) \\
2. überwiegend wird morgen wieder überall die sonne scheinen .\\
(EN: The sun will shine everywhere again tomorrow.) \\
3. es wird morgen wieder überall sonnig sein .\\
(EN: It will be sunny again everywhere tomorrow.) \\
\midrule
\textbf{Top-3 Predictions:}\\
1. \textcolor{myred}{morgen wird es wieder} \textcolor{myblue}{weitgehend} \textcolor{myred}{sonnig} \\
(EN: \textcolor{myred}{Tomorrow will be} \textcolor{myblue}{widely} \textcolor{myred}{sunny again}.) \\
2. \textcolor{myred}{es wird morgen} \textcolor{myblue}{größtenteils} \textcolor{myred}{wieder} \textcolor{myblue}{sonniges wetter geben}\\
(EN: \textcolor{myred}{Tomorrow} \textcolor{myblue}{will mostly be sunny} \textcolor{myred}{again}.) \\
3. \textcolor{myred}{morgen} \textcolor{myblue}{ist fast in der gesamten region sonnenschein} zu erwarten\\
(EN: \textcolor{myred}{Tomorrow}, \textcolor{myblue}{sunshine} is expected \textcolor{myblue}{in almost the entire region}.) \\ \midrule \midrule

\textbf{Ground Truth (CSL-Daily):} \begin{CJK}{UTF8}{gbsn}服 务 员 卫 生 间 在 哪 里\end{CJK} \\
(EN: Waiter, where is the bathroom?) \\ \midrule
\textbf{Extended Multiple References (CSL-Daily-Div):} \\ 
\begin{CJK}{UTF8}{gbsn}1. 请 问 服 务 人 员 我 可 以 在 哪 里 找 到 卫 生 间 \end{CJK} \\
(EN: Please ask the service staff where I can find the bathroom?) \\
\begin{CJK}{UTF8}{gbsn}2. 厕 所 在 哪 儿 员 工 能 告 诉 我 吗\end{CJK} \\
(EN: Where is the toilet, can the staff tell me?) \\
\begin{CJK}{UTF8}{gbsn}3. 服 务 员 你 能 告 诉 我 卫 生 间 在 哪 里 吗\end{CJK} \\
(EN: Waiter, could you tell me where the bathroom is?) \\
\midrule
\textbf{Top-3 Predictions:} \\  
\begin{CJK}{UTF8}{gbsn}1. \textcolor{myred}{服 务 员 请 问} 你 们 的 \textcolor{myred}{卫 生 间 在 哪 里}\end{CJK} \\
(EN: \textcolor{myred}{Waiter, could you please tell me where} your \textcolor{myred}{bathroom is}?) \\
    \begin{CJK}{UTF8}{gbsn}2. \textcolor{myred}{员 工 能 告 诉 我 厕 所 在 哪 吗}\end{CJK} \\
(EN: \textcolor{myred}{Can the staff tell me where the toilet is}?) \\ 
\begin{CJK}{UTF8}{gbsn}3. \textcolor{myred}{服 务 员 请 问} 你 们 的 \textcolor{myred}{厕 所} \textcolor{myblue}{在 何 处}\end{CJK} \\
(EN: \textcolor{myred}{Waiter, please tell me} \textcolor{myblue}{where is} your \textcolor{myred}{toilet}?) \\ \midrule \midrule

\textbf{Ground Truth (CSL-Daily):} \begin{CJK}{UTF8}{gbsn}如 果 我 们 不 爱 护 地 球 那 么 地 球 也 会 惩 罚 我 们\end{CJK} \\ 
(EN: If we do not take care of the Earth, then the Earth will also punish us.) \\ \midrule
\textbf{extended Multiple References (CSL-Daily-Div):} \\ 
\begin{CJK}{UTF8}{gbsn}1. 假 如 我 们 对 地 球 的 保 护 意 识 缺 失 那 地 球 也 会 惩 罚 我 们\end{CJK} \\
(EN: If we lack awareness in protecting the Earth, then the Earth will punish us.) \\
\begin{CJK}{UTF8}{gbsn}2. 假 如 我 们 不 去 爱 护 地 球 那 么 地 球 自 然 也 会 对 我 们 进 行 惩 罚\end{CJK} \\
(EN: If we do not cherish the Earth, then Earth will naturally punish us.) \\
\begin{CJK}{UTF8}{gbsn}3. 如 果 我 们 不 去 保 护 地 球 那 地 球 将 会 对 我 们 进 行 报 复\end{CJK} \\
(EN: If we do not protect the Earth, then the Earth will retaliate against us.) \\
\midrule
\textbf{Top-3 Predictions:} \\  
\begin{CJK}{UTF8}{gbsn}1. \textcolor{myred}{如 果 我 们 不 爱 护 地 球 那 么 地 球} 必 然 \textcolor{myred}{会 惩 罚 我 们}\end{CJK} \\
(EN: \textcolor{myred}{If we do not care for the Earth, then it will} inevitably \textcolor{myred}{punish us}.) \\
\begin{CJK}{UTF8}{gbsn}2. \textcolor{myblue}{假 设} \textcolor{myred}{我 们 不 爱 护 地 球 那 么 地 球} 必 \textcolor{myred}{会 给 我 们 应 有 的 惩 罚}\end{CJK} \\
(EN: \textcolor{myblue}{If we assume} \textcolor{myred}{we do not care for the Earth, then it will} certainly \textcolor{myred}{give us the punishment we deserve}.) \\
\begin{CJK}{UTF8}{gbsn}3. \textcolor{myred}{假 如 我 们 不} \textcolor{myblue}{关 爱} \textcolor{myred}{地 球 那 么 地 球} 定 \textcolor{myred}{会 对 我 们 进 行 惩 罚} \end{CJK} \\
(EN:  \textcolor{myred}{If we do not} \textcolor{myblue}{care} \textcolor{myred}{the Earth, then the Earth will} definitely \textcolor{myred}{punish us}.) \\ \midrule \midrule

\textbf{Ground Truth (CSL-Daily):} \begin{CJK}{UTF8}{gbsn}对 待 学 习 较 差 的 同 学 我 们 肯 定 要 帮 助 他\end{CJK} \\
(EN: We must certainly help students who are struggling with their studies.) \\ \midrule
\textbf{extended Multiple References (CSL-Daily-Div):} \\ 
\begin{CJK}{UTF8}{gbsn}1. 我 们 绝 对 有 义 务 帮 助 学 习 成 绩 不 佳 的 同 学 \end{CJK} \\
(EN: We absolutely have an obligation to assist students who are performing poorly academically.) \\
\begin{CJK}{UTF8}{gbsn}2. 对 于 学 习 上 有 困 难 的 同 学 我 们 无 疑 应 该 伸 出 一 只 援 手\end{CJK} \\
(EN: For students facing difficulties in their studies, we should undoubtedly extend a helping hand.) \\
\begin{CJK}{UTF8}{gbsn}3. 有 问 题 的 同 学 在 学 业 上 我 们 是 肯 定 要 给 予 援 助 的\end{CJK} \\
(EN: We are definitely committed to providing assistance to students who encounter problems in their academic pursuits.) \\
\midrule
\textbf{Top-3 Predictions:} \\  
\begin{CJK}{UTF8}{gbsn}1. \textcolor{myred}{学 习}  \textcolor{myblue}{不 好} 是 我 们 \textcolor{myred}{同 学 我 们} 应 该 \textcolor{myred}{帮 助 他} 们\end{CJK} \\
(EN: \textcolor{myred}{Assisting} our \textcolor{myred}{classmates} who are  \textcolor{myblue}{struggling academically} is something \textcolor{myred}{we} should \textcolor{myred}{undertake}.) \\
    \begin{CJK}{UTF8}{gbsn}2. \textcolor{myred}{同 学} 对 \textcolor{myred}{学 习}  \textcolor{myblue}{不 好} 
 \textcolor{myred}{ 我 们} 应 该 \textcolor{myred}{帮 助 他} 们\end{CJK} \\
(EN: \textcolor{myred}{We} should \textcolor{myred}{help our classmates} who are  \textcolor{myblue}{struggling} with their \textcolor{myred}{studies}.) \\ 
\begin{CJK}{UTF8}{gbsn}3. \textcolor{myred}{我 们} \textcolor{myblue}{必 须 对 学 习}  \textcolor{myblue}{不 好} \textcolor{myred}{同 学 帮 助}\end{CJK} \\
(EN: \textcolor{myred}{We} \textcolor{myblue}{must} \textcolor{myred}{assist classmates} who are  \textcolor{myblue}{not doing well} in their \textcolor{myred}{studies}.) \\ 
\bottomrule
\end{tabular}
\end{table*}


\end{document}